%% file: main.tex
\documentclass[]{spie}  

\usepackage{amsmath,amsfonts,amssymb}
\usepackage{caption}
\usepackage{subcaption}
\usepackage{graphicx}
\usepackage[colorlinks=true, allcolors=blue]{hyperref}
\usepackage[noabbrev]{cleveref}
\creflabelformat{equation}{#2\textup{#1}#3}
\usepackage[draft]{todonotes}
\usepackage{listings}
\usepackage{upgreek}

\title{Full characterization of the instrumental polarization effects of the spectropolarimetric mode of SCExAO-CHARIS}

\input{authors}

\begin{document} 
\maketitle

\thanks{\noindent Based  on data collected at Subaru Telescope, which is operated by the National Astronomical Observatory of Japan.}

\begin{abstract}
\input{abstract}
\end{abstract}


\section{Introduction}\label{cha:Introduction}
\input{introduction}

\section{Description of polarized light}\label{cha:Theory}
\input{theory}

\section{Optical path and instrumental polarization effects}\label{cha:Instrument}
\input{instrument}

\section{Mathematical description of optical system}\label{cha:Model}
\input{model}

\section{Calibration using internal source measurements}\label{cha:InternalSource}
\input{internal_source}

\section{Calibration using unpolarized standard star observations}\label{cha:OnSky}
\input{onsky}

\section{Polarimetric accuracy of the model}\label{cha:Accuracy}\label{sec:est_pol_acc}
\input{accuracy}

\section{Recommendations}\label{cha:Discussion}
\input{discussion}

\section{Conclusions}\label{cha:Conclusion}
\input{conclusion}

\acknowledgments 
\input{acknowledgments}

\bibliography{main} 
\bibliographystyle{spiebib} 

\appendix

\newpage

\section{Estimating the diattenuation of the derotator}\label{cha:appendix_der_diatt}
\input{appendix_diattenuation_of_derotator}

\end{document}

%% file: authors.tex
\author[a]{G. J. Joost `t Hart}
\author[a]{Rob G. van Holstein}
\author[a]{Steven P. Bos}
\author[a]{Jasper Ruigrok}
\author[a]{Frans Snik }
\author[b]{Julien Lozi}
\author[b,c,d,e]{Olivier Guyon}
\author[b,e]{Tomoyuki Kudo}
\author[f]{Jin Zhang}
\author[g]{Nemanja Jovanovic}
\author[h]{Barnaby Norris}
\author[h]{Marc-Antoine Martinod}
\author[i]{Tyler D. Groff}
\author[j]{Jeffrey Chilcote}
\author[b,q,r]{Thayne Currie}
\author[e,f,k]{Motohide Tamura}
\author[b,e,l]{S\'ebastien Vievard}
\author[b,m,n]{Ananya Sahoo}
\author[b]{Vincent Deo}
\author[b]{Kyohoon Ahn}
\author[o]{Frantz Martinache}
\author[p]{Jeremy Kasdin}

\affil[a]{Leiden Observatory, Leiden University, PO Box 9513, 2300 RA Leiden, The Netherlands}
\affil[b]{Subaru Telescope, National Astronomical Observatory of Japan, National Institutes of Natural Sciences (NINS), 650 North Aohoku Place, Hilo, HI, 96720, U.S.A.}
\affil[c]{Steward Observatory, University of Arizona, Tucson, AZ, 85721, U.S.A.}
\affil[d]{College of Optical Sciences, University of Arizona, Tucson, AZ 85721, U.S.A.}
\affil[e]{Astrobiology Center of NINS, 2-21-1, Osawa, Mitaka, Tokyo, 181-8588, Japan}
\affil[f]{University of Tokyo, 7 Chome-3-1 Hongo, Bunkyo City, Tokyo 113-8654, Japan}
\affil[g]{California Institute of Technology, 1200 E California Blvd, Pasadena, CA 91125, U.S.A.}
\affil[h]{Sydney Institute for Astronomy, Institute for Photonics and Optical Science, School of Physics, University of Sydney, NSW 2006, Australia}
\affil[i]{Goddard Space Flight Center, 8800 Greenbelt Rd, Greenbelt, MD 20771, U.S.A}
\affil[j]{Univeristy of Notre Dame, Notre Dame, IN 46556, U.S.A.}
\affil[k]{National Astronomical Observatory of Japan, National Institutes of Natural Sciences (NINS), 2 Chome-21-1 Osawa, Mitaka, Tokyo 181-0015, Japan}
\affil[l]{Observatoire de Paris, LESIA, 5 Place Jules Janssen, 92190 Meudon, France}
\affil[m]{Sokendai, Graduate University for Advanced Studies, Kanagawa Prefecture, Miura District, Hayama, Shonankokusaimura, 240-0193, Japan}
\affil[n]{Space Telescope Science Institute, 3700 San Martin Drive, Baltimore, MD 21218, USA}
\affil[o]{Laboratoire Lagrange, Universite Cote d’Azur, Observatoire de la Cote d’Azur, CNRS, Parc Valrose, Bat. H. FIZEAU, 06108 Nice, France}
\affil[p]{Princeton University, Princeton, NJ 08544, U.S.A. }
\affil[q]{NASA-Ames Research Center, Moffett Blvd., Moffett Field, CA, USA}
\affil[r]{Eureka Scientific, 2452 Delmer Street Suite 100, Oakland, CA, USA}

\authorinfo{Further author information: (Send correspondence to G.J.J. `t Hart) \\ G.J.J. `t Hart: E-mail: hart@strw.leidenuniv.nl}

%% file: abstract.tex
SCExAO at the Subaru telescope is a visible and near-infrared high-contrast imaging instrument employing extreme adaptive optics and coronagraphy. The instrument feeds the near-infrared light (JHK) to the integral-field spectrograph CHARIS. The spectropolarimetric capability of CHARIS is enabled by a Wollaston prism and is unique among high-contrast imagers. We present a detailed Mueller matrix model describing the instrumental polarization effects of the complete optical path, thus the telescope and instrument. From measurements with the internal light source, we find that the image derotator (K-mirror) produces strongly wavelength-dependent crosstalk, in the worst case converting ${\sim}95\%$ of the incident linear polarization to circularly polarized light that cannot be measured. Observations of an unpolarized star show that the magnitude of the instrumental polarization of the telescope varies with wavelength between 0.5\% and 1\%, and that its angle is exactly equal to the altitude angle of the telescope. Using physical models of the fold mirror of the telescope, the half-wave plate, and the derotator, we simultaneously fit the instrumental polarization effects in the 22 wavelength bins. Over the full wavelength range, our model currently reaches a total polarimetric accuracy between 0.08\% and 0.24\% in the degree of linear polarization. We propose additional calibration measurements to improve the polarimetric accuracy to ${<}0.1\%$ and plan to integrate the complete Mueller matrix model into the existing CHARIS post-processing pipeline. Our calibrations of CHARIS’ spectropolarimetric mode will enable unique quantitative polarimetric studies of circumstellar disks and planetary and brown dwarf companions.

%% file: introduction.tex

With near-infrared (NIR) spectropolarimetric observations, the spectra of polarized scattered light from dust grains in circumstellar disk and disks around brown dwarf companions and planets can be measured. These observations can reveal key information on the dust grain properties in these disks and can teach us about the formation of planetary systems\cite{murakawa2010dustproperties, canovas2015dustproperties, marley2011modelpolarizationexoplanet, stolker2017modelpolarizationexoplanet}. The degree of linear polarization of circumstellar disks is generally ${>}10\%$, which enables us to measure the polarized scattering phase function and derive properties of dust grains in the disk through radiative transfer modeling\cite{perrin2015GPIpolarimetry}. The degree of linear polarization of the thermal NIR radiation of a planet, scattered in its atmosphere, can be a few percent and depends strongly on the properties of its atmosphere, for instance, the temperature gradient, scattering properties, and distribution of its cloud particles\cite{deKok2011dolpsimualtions}. Furthermore, NIR light scattered from a dust disk around a substellar companion can make the companion polarized on the order of a few percent\cite{stolker2017modelpolarizationexoplanet}.


Using polarimetric differential imaging (PDI), the NIR polarimetric modes of the high-contrast imaging instruments SPHERE-IRDIS\cite{beuzit2019sphere,dohlen2008irdis,deBoer2020irdis,vanHolstein2020IRDIS} at the Very Large Telescope, Germini Planetary Imager\cite{macintosh2014GPI,perrin2015GPIpolarimetry} at the Gemini South Telescope, and HiCIAO\cite{hodapp2008hiciao} at the Subaru telescope have been able to successfully image circumstellar disk of various ages\cite{garufi20173yearsSPHERE,avenhaus2018disksSPHER,esposito2020disksGPI,hashimoto2011diskHiCIAO,muto2012diskHiCIAO}. SPHERE-IRDIS and the Gemini Planet imager have also been used to search for polarization signals from substellar and planetary objects \cite{millar2015beta, jensen2016point, van2017combining, jensen2020search, vanHolstein2021}, leading to the polarimetric detections of two spatially unresolved circumsubstellar disks\cite{vanHolstein2021}. 

In 2017, the HiCIAO instrument at the Subaru telescope was decommissioned and replaced with the Subaru Coronagraphic Extreme Adaptive Optics (SCExAO) system\cite{jovanovic2015scexao}. 
SCExAO is a multi-band instrument that operates at wavelengths from 600 to 2500 nm. It is equipped with multiple coronagraphs. A wavefront sensing and control system is key to the operation of SCExAO. A first low-order wavefront correction is performed by Subaru's facility adaptive optics (AO188)\cite{minowa2010AO188}. Further downstream in SCExAO, a second wavefront correction is performed for high-order modes using a combination of visible pyramid wavefront sensors and a 2000-element deformable mirror operating at a wavelength range of 600-900 nm. Initially, SCExAO had no NIR polarimetric capabilities. Recently, a spectropolarimetric mode has been implemented in SCExAO for its Coronagraphic High Angular Resolution Imaging Spectrograph (CHARIS) subsystem\cite{groff2017,lozi2018scexao}. CHARIS is an integral-field spectrograph that provides low-resolution spectra over the JHK-bands. The detector has a field-of-view of $2" \times 2"$. To enable spectropolarimetry to be performed, a Wollaston prism, which spatially splits the light into its two orthogonal linear polarization states, has been installed directly upstream of CHARIS. The implementation of the Wollaston prism reduces the FOV to $2" \times 1"$, because it images the orthogonal polarization states on the same detector, without the loss of spatial resolution. The spectropolarimetric mode is unique among high-contrast imagers, allowing to observe polarization signal over multiple wavelength bins simultaneously.


The accuracy of CHARIS' spectropolarimetric mode is currently limited by instrumental polarization effects of the telescope and the optical components of the instrument. Due to the instrumental polarization effects, the polarization signal measured at the detector is different from the signal incident on the telescope. The two most important effects are instrumental polarization (IP), which is the polarization signal produced by the optical components in the system, and crosstalk, the instrument-induced mixing of polarization states, which most often is the conversion of linear polarization to circular polarization and vice versa. IP can make unpolarized sources appear to be polarized. For light with a low degree of linear polarization, IP can result in an apparent rotation of the orientation of the linear polarization. Crosstalk can also cause rotation of the polarization state. Furthermore, with the Wollaston prism, the system cannot measure circular polarization, so that crosstalk from linear to circular polarization leads to a reduction of the polarimetric efficiency, that is, the fraction of the incident linear polarization that is actually measured. 
It has already been shown that the image derotator (K-mirror) in SCExAO produces strongly wavelength-dependent crosstalk, in the worst-case converting 95\% of the incident linearly polarized light into circularly polarized light\cite{vanHolstein2020calibration}.

%


To achieve a high polarimetric accuracy, we need to characterize the IP and crosstalk of the instrument, so that observations can be corrected for these instrumental polarization effects. This will enable us to measure the polarization of circumstellar disks and substellar companions with a high accuracy. Together with the spectroscopic capabilities of CHARIS, high polarimetric accuracy will enable unique research of the composition of circum(sub)stellar disks. 


In this study, we continue the work by Ref.~\citenum{vanHolstein2020calibration} and characterize the instrumental polarization effects of the entire optical system of SCExAO-CHARIS. We use a Mueller matrix model to describe the instrumental polarization effects of all relevant optical components. The parameters describing the model are determined using measurements with SCExAO's internal light source and observations of an unpolarized standard star. This method has been used successfully to calibrate the instrumental polarization effects of SPHERE-IRDIS\cite{vanHolstein2020IRDIS}. Ref.~\citenum{vanHolstein2020calibration} calibrated the retardance and diattenuation of the image derotator and half-wave plate (HWP) by fitting the retardance directly for all wavelengths separately. In this work, we use physical models to model the retardance of the image derotator and HWP and the diattenuation of the telescope mirrors. These models are fitted using all wavelength bins simultaneously.

The goal of this research is to achieve, in all wavelength bins, a polarimetric accuracy of ${<}0.1\%$ in the degree of linear polarization and an accuracy of a few degrees in the angle of linear polarization when observing a $1\%$ linearly polarized companion. These accuracies are similar to those of the Mueller matrix model developed for SPHERE-IRDIS\cite{vanHolstein2020IRDIS}. With a polarimetric accuracy of ${<}0.1\%$, we will be able to measure the polarized spectrum of substellar and planetary companions. Furthermore, these accuracies suffice for quantitative polarimetric observations of circumstellar disks\cite{vanHolstein2021,perrin2009diskdetectionexample}.

The outline of this work is as follows. In \cref{cha:Theory}, we give a brief description of polarized light. We then discuss the optical path of SCExAO-CHARIS as well as the expected instrumental polarization effects in \cref{cha:Instrument}. Subsequently, we outline the mathematical model to describe the instrumental polarization effects in \cref{cha:Model}. Next, in \cref{cha:InternalSource,cha:OnSky}, we discuss the calibrations using the internal source and on-sky measurements, respectively. After that, we estimate the accuracy of our model in \cref{cha:Accuracy} and provide recommendations for future research in \cref{cha:Discussion}. Finally, we present conclusions in \cref{cha:Conclusion}.

%% file: theory.tex
\begin{figure}[h]
    \centering
    \includegraphics[width=0.5\textwidth]{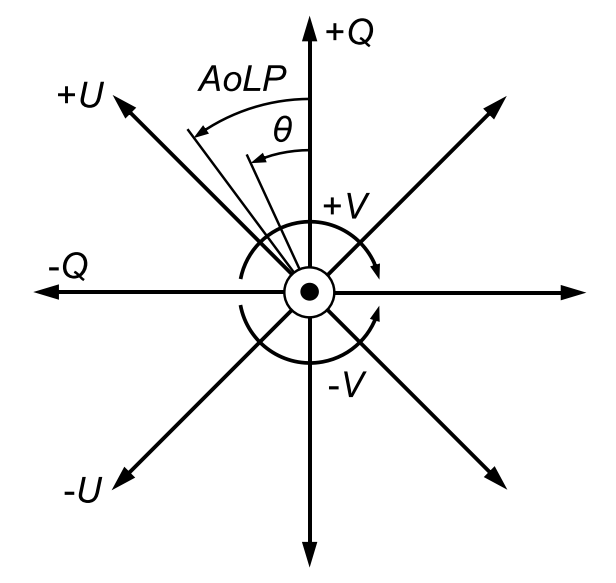}
    \caption{Definition of the reference frame of Stokes parameters and angle of linear polarization ($AoLP$) describing light propagating out of the paper, toward the reader. All rotations ($\theta$) in this work are according to this figure, positive in the counter-clockwise direction when looking into the beam of light. Figure adopted from Ref.~\citenum{vanHolstein2020IRDIS}.}
    \label{fig:stokes_convention}
\end{figure}

The polarization state of light can be described using the Stokes vector, which is defined as:
\begin{equation}
    \boldsymbol{S} =
    \begin{pmatrix}
        I \\ 
        Q \\
        U \\
        V
    \end{pmatrix},
\end{equation}
where $I$ is the total intensity of the light, $Q$ and $U$ describe the two orthogonal linear polarization states, and $V$ describes the amount and handedness of circular polarization\cite{born2013principles}. \Cref{fig:stokes_convention} shows the conventions of the Stokes parameters used throughout this work. In this figure, the positive $z$ direction is oriented out of the paper toward the reader. These conventions are the same conventions as used by Ref.~\citenum{vanHolstein2020IRDIS}.

The Stokes vector can be normalized by dividing each of the Stokes parameters by the total intensity $I$:
\begin{equation}
    \boldsymbol{S} =
    \begin{pmatrix}
        1 \\ 
        q \\
        u \\
        v
    \end{pmatrix},
\end{equation}
where $q$, $u$, and $v$ are the normalized Stokes parameters.
Using these quantities, the degree of linear polarization $P$ and angle of linear polarization $\chi$ can be calculated as:
\begin{align}
    \label{eq:dolp}
    P &= \sqrt{q^2 + u^2}, \\
    \label{eq:aolp}
    \chi &= \frac{1}{2} \arctan \left( \frac{u}{q} \right).
\end{align}

%% file: instrument.tex
With the conventions in place, we can now look at the instrument, SCExAO-CHARIS. This section contains two parts. First, we discuss the optical path of SCExAO-CHARIS in \cref{sec:optical_path}. Subsequently, we discuss the instrumental polarization effects of the components in \cref{sec:polarization_effects_of_instrument}.

\subsection{Optical Path}\label{sec:optical_path}
\begin{figure}[h]
    \centering
    \includegraphics[width=\textwidth]{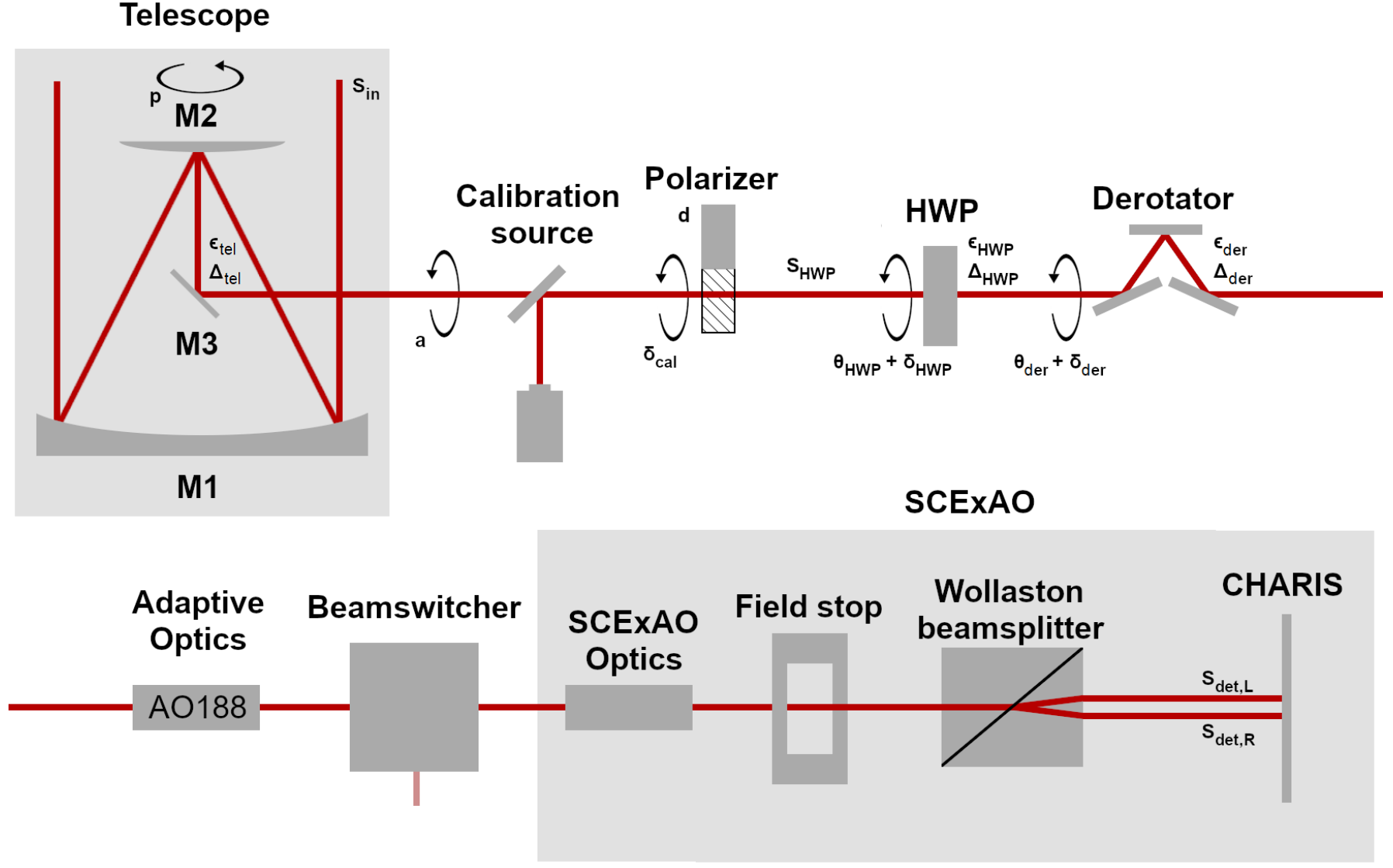}
    \caption{Overview of the optical path showing only components relevant for polarimetric measurements. The names of the components are given in bold font. The black circular arrows show the parallactic, altitude, and component rotations about the optical axis. The relevant parameters for calibrating the instrumental polarization effects are shown next to the components. For on-sky observations, the calibration source and polarizer are moved out of the beam. Figure adopted from Ref.~\citenum{vanHolstein2020calibration}.}
    \label{fig:optical_path}
\end{figure}

\Cref{fig:optical_path} shows a simplified schematic of the optical path of the Subaru telescope, AO188, and SCExAO-CHARIS, containing all components relevant for spectropolarimetric observations. The Subaru telescope is located at the summit of Mauna Kea at 4.2 km altitude. During observations, the light is collected by the 8-m primary mirror (M1). The light from M1 is then reflected by the secondary mirror (M2) that is suspended at the top of the telescope. This light subsequently hits a flat tertiary mirror (M3), which reflects the light to the Nasmyth platform where SCExAO is installed. Downstream of the telescope, an insertable and rotatable broadband half-wave plate (HWP) is located. 
The HWP is used to rotate the angle of linear polarization of the incoming light, thereby selecting the incident polarization state to be measured. After the light passes the HWP it reaches an image derotator, which is a construction of three mirrors (K-mirror) at incidence angles of $60^\circ$, $30^\circ$, and $60^\circ$, respectively. The derotator rotates in order to compensate for the sky and telescope rotation during the observations. The derotator is part of the adaptive optics system AO188\cite{minowa2010AO188}.  All reflections in AO188 are in the horizontal plane (i.e.,~parallel to the Nasmyth platform). In the future, a beam switcher will be installed after the adaptive optics system, which will direct the light to the different instruments on the platform\cite{lozi2020beamswitcher}.

SCExAO is located directly downstream of AO188. SCExAO has several coronagraphs, but we expect none of them to produce significant instrumental polarization effects. Inside SCExAO, the light undergoes multiple reflections in the horizontal plane, after which it passes a field stop. Next, the light passes a Wollaston prism that spatially splits the light into the vertical and horizontal linear polarization states. The light then passes through a lenslet array and a prism that create the polarized spectra of the source on the Hawaii2RG detector. The vertical and horizontal polarization states are recorded on the right and left side of the detector, respectively.

For calibration measurements, a system that can be inserted into the optical path to perform internal calibration measurements is located directly downstream of the telescope on the Nasmyth platform. The system contains an internal light source, a flat mirror, and a rotatable wire-grid polarizer, which together produce linearly polarized light.

From the measured intensity on the right and left sides of the detector, $I_\mathrm{det, R}$ and $I_\mathrm{det, L}$, Stokes $Q$ and $U$ and their corresponding total intensities, $I_Q$ and $I_U$, can be determined using the single difference and single sum, respectively, as:
\begin{gather}
    \label{eq:single_dif}
    X^\pm = I_\mathrm{det, R} - I_\mathrm{det, L}, \\
    \label{eq:single_sum}
    I_{X^\pm} = I_\mathrm{det, R} + I_\mathrm{det, L},
\end{gather}
where $X^\pm$ is the single difference and $I_{X^\pm}$ the single sum of Stokes $Q$ or $U$. $Q$ is measured with the HWP at $0^\circ$ and $U$ with the HWP at $22.5^\circ$. The resulting single differences are called $Q^+$ and $U^+$ and the single sums $I_{Q^+}$ and $I_{U^+}$. Note that we subtract the intensity of the left detector side from the intensity of the right detector side. This is opposite to SPHERE-IRDIS\cite{vanHolstein2020IRDIS} and the definitions assumed in Ref.~\citenum{vanHolstein2020calibration}. We can additionally measure Stokes $Q$ and $U$ with the HWP at $45^\circ$ and $67.5^\circ$, respectively. The results of these measurements are called $Q^-$, $U^-$, $I_{Q^-}$, and $I_{U^-}$. One set of observations with HWP angles rotated by $0^\circ$, $45^\circ$, $22.5^\circ$, $67.5^\circ$ is called a HWP cycle. For a perfect optical system $X^+$ and $X^-$, and $I_{X^+}$ and $I_{X^-}$ will be identical and of opposite sign. However, due to instrumental polarization effects and varying atmospheric seeing the measurements differ.

\subsection{Instrumental Polarization Effects of Optical Components}\label{sec:polarization_effects_of_instrument}
All optical components that reflect or transmit light produce IP and crosstalk. IP is a result of the diattenuation of an optical component. Diattenuation is caused by the difference in reflectance and transmittance of orthogonal linear polarization states. For instance, the reflectance of light polarized perpendicular to the plane of incidence of a mirror is generally higher than the reflectance of light polarized parallel to this plane\cite{breckinridge2015polarization}. This results in the creation of a polarized signal by the mirror. Crosstalk is the result of the retardance of the component, that is, the relative phase shift of the different linear polarization states after reflection or transmission. Diattenuation and retardance depend on the optical properties of a component, which are generally a function of wavelength, as well as the angle of incidence of the light onto the optical component.

The diattenuation and retardance of mirrors are generally strongest for reflections at large angles of incidence. The largest effects are thus expected from the reflections of M3 and the derotator mirrors. The diattenuation and retardance of M1 and M2 are expected to be small because these mirrors are rationally symmetric with respect to the optical axis. The reflections in AO188 are all at small angles of incidence and therefore the diattenuation and retardance are expected to be small. Due to imperfections in the design and manufacturing of the HWP, the retardance is expected to differ slightly from the ideal value of $180^\circ$, leading to a small amount of crosstalk.

The IP of all stationary components downstream of the HWP can be removed using beam switching with the HWP. As discussed in \cref{sec:optical_path}, Stokes $Q$ can be measured with the HWP oriented at $0^\circ$ and $45^\circ$. By rotating the HWP over $45^\circ$, the sign of the measured polarization signal changes, but the sign of the IP created downstream of the HWP remains the same (see e.g., Ref.~\citenum{deBoer2020irdis}). For an imperfect polarimeter we thus measure in one HWP cycle $Q^+=Q + \mathrm{IP}$ and $Q^- = -Q + \mathrm{IP}$, and similar for observations of Stokes $U$. The observations with the HWP switched by $45^\circ$ can then be combined using the double difference and double sum, as:
\begin{gather}
    \label{eq:double_diff}
    X = \frac{1}{2}\left( X^+ - X^-\right), \\
    \label{eq:double_sum}
    I_X = \frac{1}{2}\left( I_{X^+} + I_{X^-} \right),
\end{gather}
where $X$ is the double-difference $Q$ or $U$, $I_X$ the double-sum $I_Q$ or $I_U$, and $X^\pm$ and $I_{X^\pm}$ the single difference and single sum calculated with \cref{eq:single_dif,eq:single_sum}, respectively. Apart from correcting for IP, the double-difference method also corrects for flat-fielding errors and differential aberrations\cite{tinbergen2005doubledifference,canovas2011datareductionpolarimetry,vanHolstein2020IRDIS}. The normalized Stokes parameters can then be calculated from the double difference and double sum as:
\begin{equation}
    \label{eq:norm_stokes}
    x = \frac{X}{I_X},
\end{equation}
where $x$ is the normalized stokes parameter $q$ or $u$.

Because all reflections downstream of the derotator lie in the horizontal plane, these reflections can only produce crosstalk between linearly polarized light oriented at $\pm45^\circ$ with respect to the horizontal plane and circularly polarized light. Light that is polarized vertically or horizontally is not affected by crosstalk. Because the Wollaston prism only splits light into its horizontal and vertical polarization states, no crosstalk of components downstream the derotator can affect the measurements. The remaining IP and crosstalk effects are due to the optics upstream of the HWP and the optics that are rotating during observations.

%% file: model.tex
In \cref{cha:Instrument}, we have identified the components that produce the most significant IP and crosstalk that are not corrected by the double difference. In this section, we give a mathematical description of the relevant components. We first introduce the Mueller matrix model and a description of the entire instrument in terms of Mueller matrices in \cref{sec:mueller_matrix_model}. In \cref{sec:derotator_model,sec:hwp_model,sec:m3_model} we then present physical models of the retardance of the derotator, the retardance of the HWP, and the diattenuation of M3, respectively.

\subsection{Mueller Matrix Model}\label{sec:mueller_matrix_model}
We use a Mueller matrix model to mathematically describe all components in the optical path of SCExAO-CHARIS. This Mueller matrix model is similar to the model of SPHERE-IRDIS by van Ref.~\citenum{vanHolstein2020IRDIS}. Consider an incident beam of light described by a Stokes vector, $\boldsymbol{S}$. If this beam of light interacts with an optical component, its polarization and/or intensity (and hence, its Stokes vector) changes. If these changes are linear, they can be described with a Mueller matrix\cite{tinbergen2005doubledifference}. A Mueller matrix is a $4 \times 4$ matrix and can be applied to a Stokes vector as follows:
\begin{equation}\label{eq:apply_mueller}
    \boldsymbol{S}_{\mathrm{out}} = M\boldsymbol{S}_\mathrm{in},
\end{equation}
where $\boldsymbol{S}_\mathrm{in}$ is the incident Stokes vector and $\boldsymbol{S}_\mathrm{out}$ the Stokes vector after the optical component. The Mueller matrix describes the transformation from one polarization state to another. The Mueller matrix transforms the Stokes vector as:
\begin{equation}\label{eq:schematic_mueller}
    \begin{bmatrix} 
        I_\mathrm{out} \\
        Q_\mathrm{out} \\ 
        U_\mathrm{out} \\ 
        V_\mathrm{out} 
    \end{bmatrix} 
    = 
    \begin{bmatrix} 
    I \hspace{-2pt}\rightarrow\hspace{-2pt} I & Q \hspace{-2pt}\rightarrow\hspace{-2pt} I & U \hspace{-2pt}\rightarrow\hspace{-2pt} I & V \hspace{-2pt}\rightarrow\hspace{-2pt} I \\
    I \hspace{-2pt}\rightarrow\hspace{-2pt} Q & Q \hspace{-2pt}\rightarrow\hspace{-2pt} Q & U \hspace{-2pt}\rightarrow\hspace{-2pt} Q & V \hspace{-2pt}\rightarrow\hspace{-2pt} Q \\
    I \hspace{-2pt}\rightarrow\hspace{-2pt} U & Q \hspace{-2pt}\rightarrow\hspace{-2pt} U & U \hspace{-2pt}\rightarrow\hspace{-2pt} U & V \hspace{-2pt}\rightarrow\hspace{-2pt} U \\
    I \hspace{-2pt}\rightarrow\hspace{-2pt} V & Q \hspace{-2pt}\rightarrow\hspace{-2pt} V & U \hspace{-2pt}\rightarrow\hspace{-2pt} V & V \hspace{-2pt}\rightarrow\hspace{-2pt} V
    \end{bmatrix} 
    \begin{bmatrix} 
        I_\mathrm{in} \\ 
        Q_\mathrm{in} \\ 
        U_\mathrm{in} \\ 
        V_\mathrm{in} 
    \end{bmatrix},
\end{equation}
where an element $A \hspace{-2pt}\rightarrow\hspace{-2pt} B$ describes how much of incident Stokes parameter $A$ is transformed into Stokes parameter $B$. The total transmission and/or reflection of the optical component is given by $I \hspace{-2pt}\rightarrow\hspace{-2pt} I$. All other elements $I \rightarrow\hspace{-2pt} X$ describe instrumental polarization, with $X$ any of the Stokes parameters $Q$, $U$ and $V$. The elements $X \rightarrow\hspace{-2pt} I$ describe instrumental depolarization, which is not expected to play a significant roll in SCExAO-CHARIS. The crosstalk of a component is described by the elements $X \rightarrow\hspace{-2pt} Y$, with $X$ and $Y$ any combination of different Stokes parameters $Q$, $U$ and $V$.

To describe the total effect of all optical components in SCExAO-CHARIS, we can expand \cref{eq:apply_mueller} as:
\begin{equation}\label{eq:mueller_entire_instrument}
    \boldsymbol{S}_\mathrm{det,L/R} = M_n M_{n-1} \dots M_2 M_1 \boldsymbol{S}_\mathrm{in},
\end{equation}
with $M_i$ the Mueller matrix of component $i$ and $S_\mathrm{det,L/R}$ the Stokes vector of the light on the left or right side of the detector. As discussed in \cref{sec:polarization_effects_of_instrument}, not all optical components have to be included in \cref{eq:mueller_entire_instrument}. Furthermore, optical components that share a fixed reference frame can be grouped together into one Mueller matrix, for example the three mirrors of the image derotator (see \cref{fig:optical_path}). Therefore, we only need to consider the following 4 component groups: $M_\mathrm{tel}$, the three telescope mirrors; $M_\mathrm{HWP}$, the half-wave plate; $M_\mathrm{der}$, the three derotator mirrors, and $M_\mathrm{BS,L/R}$, the optical path downstream the derotator, including the Wollaston prism.

To limit the number of model parameters, the Mueller matrices of the telescope, HWP, and the detector are modeled as a function of their diattenuation ($\epsilon$) and retardance ($\varDelta$)\cite{keller2002instrumentation}:
\begin{equation}\label{eq:mueller_component}
    M_\mathrm{com} = \begin{bmatrix}
        1 & \epsilon & 0 & 0 \\
        \epsilon & 1 & 0 & 0 \\
        0 & 0 &  \phantom{-}\sqrt{1-\epsilon^2}\cos\varDelta & \sqrt{1-\epsilon^2}\sin\varDelta \\
        0 & 0 & -\sqrt{1-\epsilon^2}\sin\varDelta & \sqrt{1-\epsilon^2}\cos\varDelta
    \end{bmatrix}.
\end{equation}
The elements $(0,1)$ and $(1,0)$ account for the IP generated by the optical component and the elements that account for crosstalk are $(2,2)$, $(2,3)$, $(3,2)$ and $(3,3)$. We ignore the total transmission of the components in \cref{eq:mueller_component}. The true transmission of the optical components is not important, because the normalized Stokes parameters $q$ and $u$ are always measured relative to the total intensity $I$. For the HWP, $M_\mathrm{com}$ is defined with the $+Q$-direction along one of its fast axis. For all mirrors, $M_\mathrm{com}$ is defined with the $+Q$-direction perpendicular to the plane of incidence. Ideally, $\epsilon =0$ and $\varDelta = 180^\circ$ for the telescope, HWP, and derotator.  This definition can only be used for both the HWP and telescope and derotator mirrors since each group of mirrors consists of an odd number of mirrors. Stokes $U$ and $V$ change sign after reflection from a mirror, thus resulting in an ideal retardance of $180^\circ$. For an even number of mirrors, $\varDelta=0^\circ$ would be ideal. The variables $\epsilon$ and $\varDelta$ depend on the angle of incidence and the wavelength of the light. 

By using the double-difference method, all instrumental polarization effects of stationary components after the HWP are corrected for and can be neglected in the model. Therefore, the Mueller matrix of the optical path downstream the derotator, including the Wollaston prism, reduces to:
\begin{equation}\label{eq:mueller_calibrato1r}
    M_\mathrm{BS} = \dfrac{1}{2}\begin{bmatrix}
        1 & \pm \epsilon_\mathrm{BS} & 0 & 0 \\
        \pm \epsilon_\mathrm{BS} & 1 & 0 & 0 \\
        0 & 0 & \sqrt{1-\epsilon_\mathrm{BS}^2} & 0 \\
        0 & 0 & 0 & \sqrt{1-\epsilon_\mathrm{BS}^2}
    \end{bmatrix},
\end{equation}
where $\epsilon_\mathrm{BS}$ is the diattenuation of the Wollaston prism that accounts for imperfect separation angles and extinction of orthogonal polarization states. The plus and minus signs are used for the vertical (right detector side) and horizontal (left detector side) transmission axis, respectively.

The component groups in the instrument rotate with respect to each other about the optical axis during observations. This rotation can be included by multiplying the Mueller matrices with rotation matrices of the form:
\begin{equation}\label{eq:rotation_matrix}
	T(\theta) = \begin{bmatrix} 
	    1 & 0 & 0 & 0 \\ 
	    0 & \cos(2\theta) & \sin(2\theta) & 0 \\ 
	    0 & -\sin(2\theta) & \cos(2\theta) & 0 \\ 
	    0 & 0 & 0 & 1 
    \end{bmatrix},
\end{equation}
where $\theta$ is the counter-clockwise rotation of the optical component when looking into the beam along the optical axis (see \cref{fig:stokes_convention}). In \cref{eq:mueller_entire_instrument}, a Mueller matrix of a component that has been rotated by an angle $\theta$ will become:
\begin{equation}
    M_\mathrm{com}(\theta) = T(-\theta)M_\mathrm{com}T(\theta).
\end{equation}

Now that we have defined the Mueller matrices, the complete optical system can be described as:
\begin{equation}
    \boldsymbol{S}_\mathrm{det, L/R} = M_\mathrm{BS, L/R}T(-\Theta_\mathrm{der}) M_\mathrm{der} T(\Theta_\mathrm{der}) T(-\Theta_\mathrm{HWP}) M_\mathrm{HWP} T(\Theta_\mathrm{HWP}) T(a) M_\mathrm{tel} T(p)\boldsymbol{S}_\mathrm{in}.
\end{equation}
When a target is tracked during an observation, it rotates with respect to the telescope with the parallactic angle $p$, whereas the telescope rotates with respect to the instrument with the telescope altitude angle $a$. For $a$ the convention is used that when the telescope is pointing at the horizon $a=0^\circ$ and $a=90^\circ$ when the telescope is pointing at zenith. The rotation of the HWP and derotator are $\Theta_\mathrm{HWP}$ and $\Theta_\mathrm{der}$. The misalignment of component $i$ is included in $\Theta_i$ as $\Theta_i = \theta_i + \delta_i$. With the first element of  $\boldsymbol{S}_\mathrm{det, L/R}$, we estimate the intensity on the detector sides. We use the calculated $\boldsymbol{S}_\mathrm{det, L/R}$ with \cref{eq:single_dif,eq:single_sum,eq:double_sum,eq:double_diff,eq:norm_stokes} to calculate the normalized Stokes parameters. 

\subsection{Physical Model of the Retardance of the Derotator}\label{sec:derotator_model}
\begin{figure}
    \centering
    \includegraphics[width=0.5\textwidth]{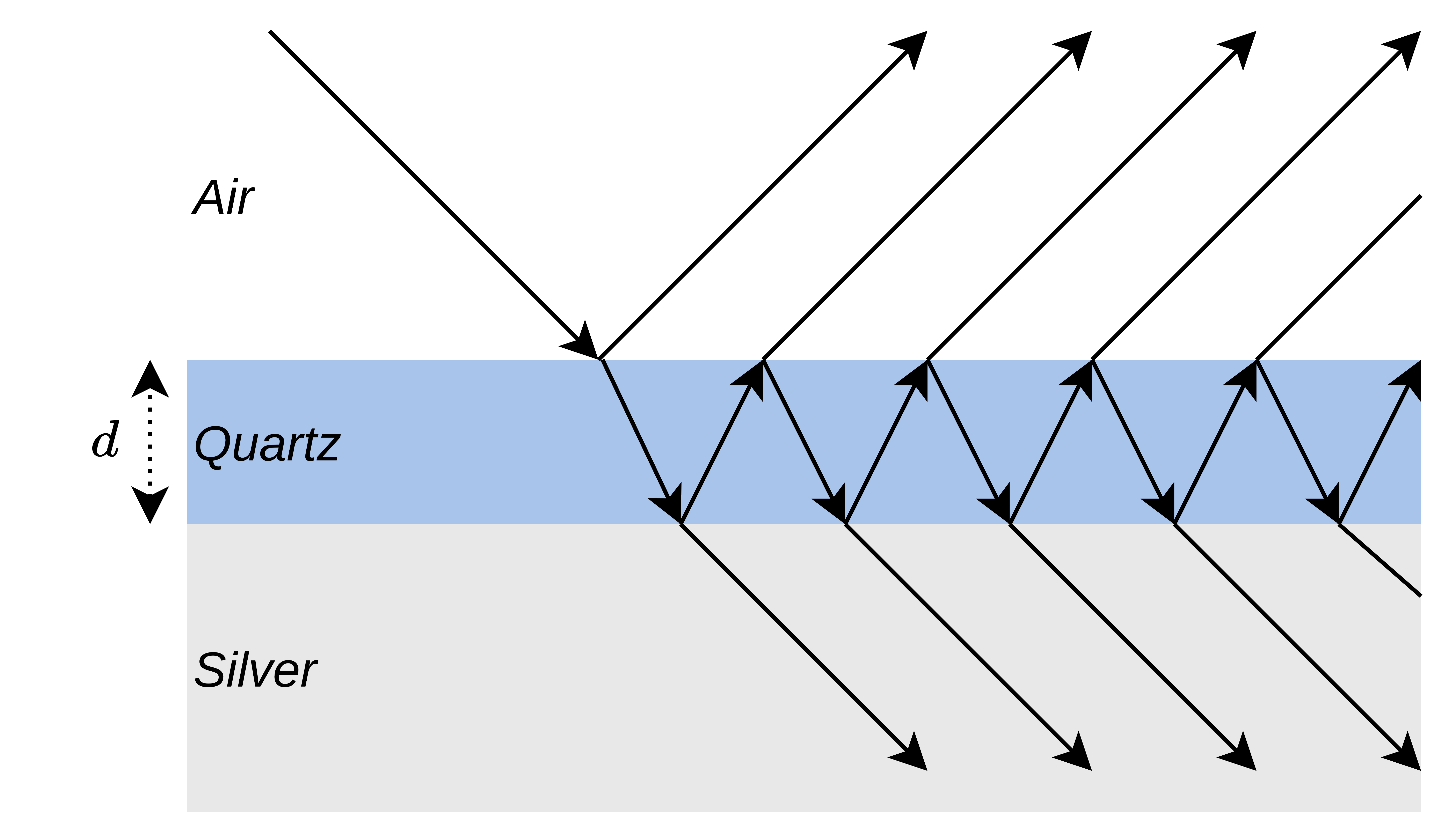}
    \caption{Conceptual drawing of the multi-layer reflection of the derotator mirrors. The width of the quartz layer is indicated as $d$.}
    \label{fig:multi_layer_reflection}
\end{figure}

The derotator in SCExAO-CHARIS consists of three mirrors with angles of incidence equal to $60^\circ$, $30^\circ$, and $60^\circ$. The mirrors are made of silver coated with quartz ($\mathrm{SiO}_2$). To accurately model the retardance of a coated mirror, multi-layer reflections have to be taken into account. The multi-layer reflection is drawn schematically in \cref{fig:multi_layer_reflection}. The incident light is reflected from and transmitted through the thin film on the silver mirror. Subsequently, the transmitted light that reaches the border between the quartz film and the silver mirror is reflected from and transmitted through the silver mirror. The light that is reflected from the silver then again reaches the edge of the thin quartz film, where part of the light is transmitted and part of the light is reflected. A part of the incident light thus bounces back and forth between the air-quartz and quartz-silver intersections. Each reflection and transmission can be described using the Fresnel equations\cite{born2013principles}. The total reflection coefficient $r_\mathrm{tot}$ is the superposition of all these reflections:
\begin{equation}\label{eq:multy-layer_reflection_infinite_sum}
    r_\mathrm{tot} = r_{01} + t_{01}r_{12}t_{10}\mathrm{e}^{\mathrm{i}\beta} + t_{01}r_{12}r_{10}r_{12}t_{10}\mathrm{e}^{2\mathrm{i}\beta}+\dots,
\end{equation}
where $r_{ij}$ and $t_{ij}$ are the reflection and transmission coefficients, respectively, from medium $i$ to medium $j$. The subscripts 0, 1, and 2 denote the media air, quartz, and silver, respectively. We assume that the reflective index of air is 1. The refractive index of quartz we obtain from Ref.~\citenum{gao2012thinfilm} and the refractive index of silver from Ref~\citenum{rakic1998optical}. \Cref{eq:multy-layer_reflection_infinite_sum} applies to both the s- and p-polarizations, which have different reflection and transmission coefficients. The phase $\beta$ accumulated by the internally reflected beam when propagating through the quartz layer is given by: 
\begin{equation}
    \beta = 2\frac{2\uppi}{\lambda} d n_1 \cos(\theta_\mathrm{t}),
\end{equation}
where $\lambda$ is the wavelength of the light, $d$ the width of the film, $n_1$ the refractive index of quartz, and $\theta_\mathrm{t}$ the angle of transmitted light in the quartz layer\cite{hecht2017optics}. The factor 2 comes from the back and forth propagation of the light through the film. Using the relation for a geometric series, $\sum_{i=0}^\infty r^i = \frac{1}{1-r}$, for a complex number $r$ with a modulus less than one, \cref{eq:multy-layer_reflection_infinite_sum} can be rewritten as\cite{lavrinenko2018numerical}:
\begin{equation}
    r_\mathrm{tot} = r_{01} + \frac{t_{01}t_{10}r_{12}\mathrm{e}^{\mathrm{i}\beta}}{1-r_{10}r_{12}\mathrm{e}^{\mathrm{i}\beta}}.
\end{equation}
The reflection coefficients of the entire derotator are the product of $r_\mathrm{tot}$ at incidence angles $60^\circ$, $30^\circ$ and $60^\circ$. Using the reflection coefficients of the entire derotator for the s- and p-polarization, the retardance of the derotator can be calculated. From the total reflection coefficients, $r_\mathrm{tot,s}$ and $r_\mathrm{tot,p}$, we calculate the retardance of the derotator as:
\begin{equation}
    \varDelta_\mathrm{der} = \arg\left(r_\mathrm{tot,s}\right) - \arg\left(r_\mathrm{tot,p}\right).
\end{equation}

\subsection{Physical Model of the Retardance of the HWP}\label{sec:hwp_model}
Wave plates are made of birefringent materials of which the refractive index is different along two perpendicular axes called the fast and slow axis. Wave plates therefore produce a phase difference between the linear polarization along the fast and slow axes of the wave plate. This retardance can be calculated as:
\begin{equation}
    \varDelta_\mathrm{HWP} = \frac{2\uppi w}{\lambda}\left[n_\mathrm{e}(\lambda) - n_\mathrm{o}(\lambda)\right],
\end{equation}
where $w$ is the thickness of the plate, $\lambda$ the wavelength of the light, and $n_\mathrm{e}$ and $n_\mathrm{o}$ are the refractive indices of the fast and slow axis, respectively. For a perfect half-wave plate $\varDelta_\mathrm{HWP}=\uppi$. Because $n_\mathrm{e}$ and $n_\mathrm{o}$ are a function of $\lambda$, achromatic wave plates are created by adding together layers of different birefringent material to create approximate constant retardance as a function of wavelength.

The achromatic HWP used in SCExAO-CHARIS consists of two layers. One layer is made of $\mathrm{SiO}_2$ crystal and the other layer is made of $\mathrm{MgF}_2$. The retardance of the achromatic HWP can be described as: 
\begin{equation}\label{eq:retardance_hwp}
    \varDelta_\mathrm{HWP} = \frac{2\uppi}{\lambda}\left[ w_{\mathrm{SiO}_2}\left[n_{\mathrm{e},\mathrm{SiO}_2}(\lambda) - n_{\mathrm{o},\mathrm{SiO}_2}(\lambda)\right] \pm w_{\mathrm{MgF}_2}\left[n_{\mathrm{e},\mathrm{MgF}_2}(\lambda) - n_{\mathrm{o},\mathrm{MgF}_2}(\lambda)\right]\right].
\end{equation}
The refractive index of $\mathrm{SiO}_2$ is obtained from Ref.~\citenum{ghosh1999SiO2} and the refractive index of $\mathrm{MgF}_2$ from Ref.~\citenum{dodge1984MgF2}. We have tested multiple documented refractive indices but found no significant differences in the resulting retardances. The $\pm$-sign denotes whether the fast axes of the layers are aligned ($+$) or the fast and slow axes are aligned ($-$). For the HWP in SCExAO, the fast and slow axes are aligned and the $-$-sign is used in \cref{eq:retardance_hwp}.

\begin{figure}[h]
    \centering
    \includegraphics[width=0.5\textwidth]{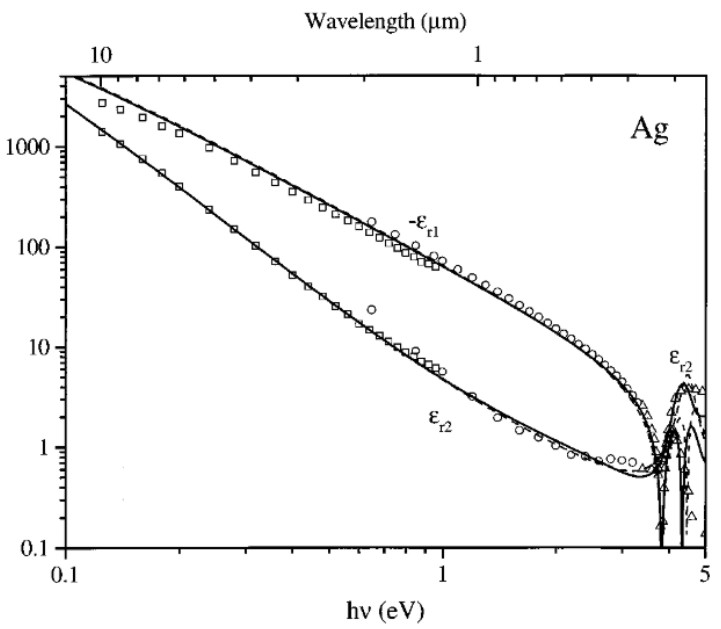}
    \caption{Real and imaginary parts of the optical dielectric functions of silver. Note that wavelength is on the upper x-axis and the corresponding photon energy is on the lower x-axis. The y-axis shows the value of the dielectric functions $\epsilon_\mathrm{r1}(\lambda)$ and $\epsilon_\mathrm{r2}(\lambda)$. Figure adopted from Ref.~\citenum{rakic1998optical}.}
    \label{fig:rakic_dielectric_functions}
\end{figure}

\subsection{Physical Model of the Diattenuation of M3}\label{sec:m3_model}
As discussed in \cref{sec:polarization_effects_of_instrument}, the instrumental polarization effects of the telescope are expected to be caused almost exclusively by the silver-coated fold mirror M3. To model the diattenuation of M3, we would preferably use the refractive index of the mirror as a function of wavelength and calculate the diattenuation using the Fresnel equations. However, the refractive index of M3 is unknown. Therefore, we create a model to describe the refractive index and diattenuation of M3.

We model the diattenuation of M3 by describing the effective refractive index of the mirrors with a linear approximation in log-log space of the dielectric functions \cite{rakic1998optical}. For metallic media, some part of the transmitted light will be absorbed. In that case, the refractive index, $n$, is a complex number of the form $n = \hat{n} + \mathrm{i}\kappa$, where $\hat{n}$ is the refractive index relating to the phase velocity in the medium, and $\kappa$ the extinction coefficient. The refractive index, $\hat{n}$, and extinction coefficient, $\kappa$, can be calculated from the dielectric functions as:
\begin{align}
    \hat{n} &= \frac{1}{\sqrt{2}} \sqrt{\sqrt{ \epsilon_\mathrm{r1}(\lambda)^2 + \epsilon_\mathrm{r2}(\lambda)^2} + \epsilon_\mathrm{r1}(\lambda)},  \\
    \kappa  &= \frac{1}{\sqrt{2}} \sqrt{\sqrt{ \epsilon_\mathrm{r1}(\lambda)^2 + \epsilon_\mathrm{r2}(\lambda)^2} - \epsilon_\mathrm{r1}(\lambda)},
\end{align}
with $\epsilon_\mathrm{r1}(\lambda)$ and $\epsilon_\mathrm{r2}(\lambda)$ the dielectric functions, which are both a function of wavelength, $\lambda$. \Cref{fig:rakic_dielectric_functions} is adopted from Ref.~\citenum{rakic1998optical} and shows the dielectric functions of silver as a function of wavelength. Assuming that for the wavelength range 1-3 $\upmu\mathrm{m}$ the dielectric functions can be approximated with a linear function in log-log space, we can write:
\begin{align}
    \label{eq:dielectric_n}
    \epsilon_\mathrm{r1}(\lambda) &= -10^{b_1}\lambda^{m_1}, \\
    \label{eq:dielectric_k}
    \epsilon_\mathrm{r2}(\lambda) &= 10^{b_2}\lambda^{m_2}
\end{align}
where $\lambda$ is the wavelength of light and $b_1$, $b_2$, $m_1$, and $m_2$ are the free parameters of the approximations. If we know the values of $b_1$, $b_2$, $m_1$, and $m_2$, we can use \cref{eq:dielectric_n,eq:dielectric_k} to compute $n$ of M3. Using $n$, we can calculate the reflection coefficients from the Fresnel equations and subsequently the diattenuation of M3, $\epsilon_\mathrm{M3}$, with:
\begin{equation}
    \epsilon_\mathrm{M3} = \frac{\left|r_\mathrm{s}\right|^2 - \left|r_\mathrm{p}\right|^2}{\left|r_\mathrm{s}\right|^2 + \left|r_\mathrm{p}\right|^2},
\end{equation}
where $r_i$ is the Fresnel reflection coefficient in the $i$ direction.

%% file: internal_source.tex
Now that we have the theoretical description of the instrument, we can estimate the model parameters by using calibration data. With the physical models of the HWP, derotator, and M3, we only need to fit seven parameters to describe the instrumental polarization effects of the entire system, namely, the width of the derotator mirror film, $d$; the widths of the HWP layers, $w_{\mathrm{SiO}_2}$ and $w_{\mathrm{MgF}_2}$; and four parameters describing the refractive index of M3, $b_1$, $b_2$, $m_1$, and $m_2$. In this section, we perform the calibration of all components downstream of the telescope mirrors. In \cref{sec:internal_source_methods}, we discuss the data set used to estimate the model parameters and the methods we use to fit the parameters. Subsequently, we present and discuss the results of the fits in \cref{sec:internal_source_results}.

\subsection{Calibration Measurements and Estimation of Model Parameters}\label{sec:internal_source_methods}
On February 18, 2020, we obtained 340 exposures with the internal calibration source. For these measurements, we inserted the calibration polarizer, generating nearly 100\% polarized light. We set the polarizer at an angle of $-45^\circ$ with respect to the instrument frame (generating a $-U$ signal). We rotated the derotator during these observations from $45^\circ$ to $127.5^\circ$ in steps of $7.5^\circ$. At each derotator orientation, we rotated the HWP from $0^\circ$ to $78.5^\circ$ with steps of $11.25^\circ$. At each derotator-HWP combination, we took multiple frames. We have preprocessed the raw data with the CHARIS data-reduction pipeline using the standard settings\cite{brandt2017data}. The data reduction produces for each observation 22 images, that is one for each wavelength bin. With this dataset we measure the retardance of the derotator and HWP ($\varDelta_\mathrm{der}$ and $\varDelta_\mathrm{HWP}$), their alignment offset angles ($\delta_\mathrm{der}$ and $\delta_\mathrm{HWP}$) and the diattenuation of the calibration polarizer ($\epsilon_\mathrm{cal}$) and its offset angle ($\delta_\mathrm{cal}$).

\begin{figure}
    \centering
    \includegraphics[width=0.5\textwidth]{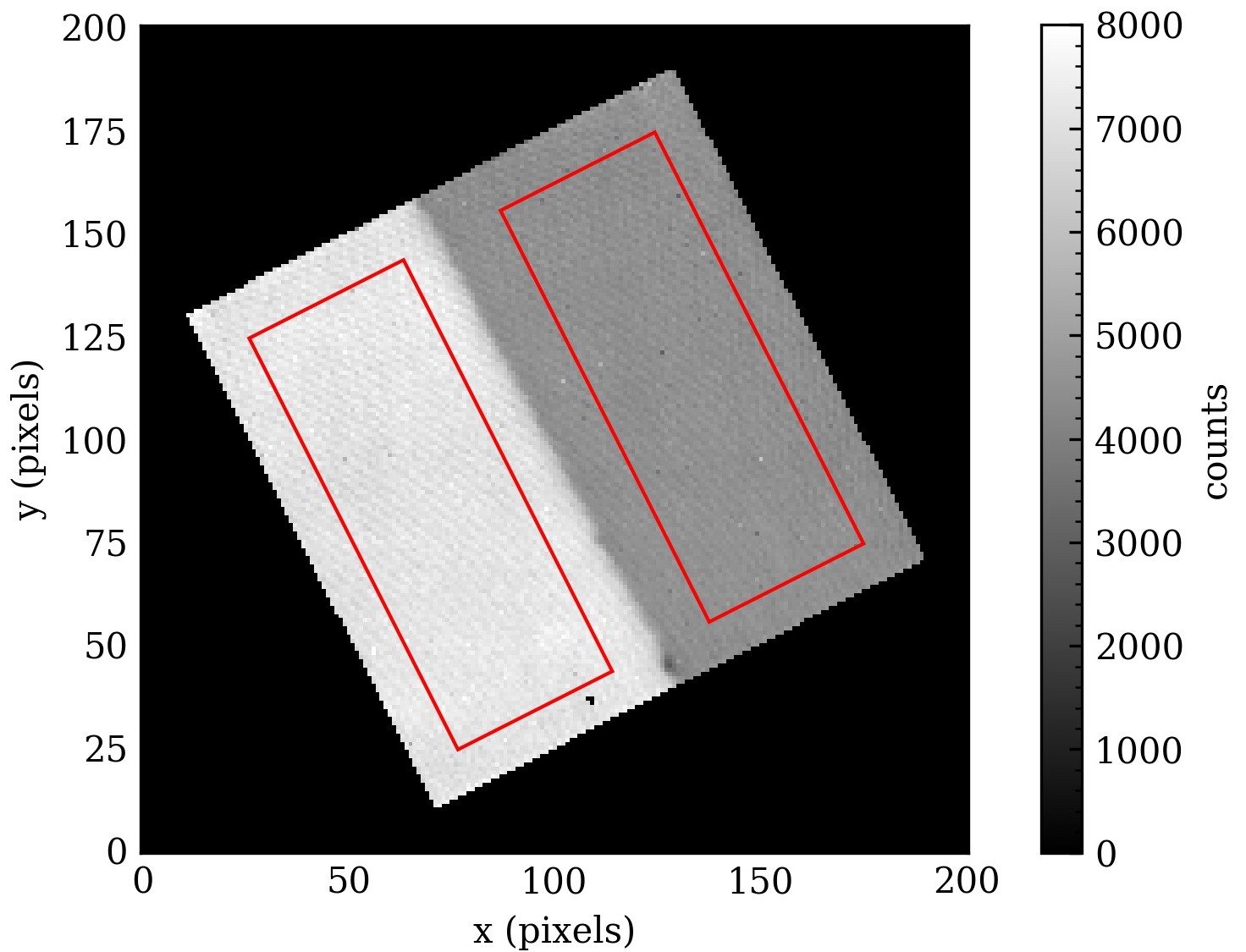}
    \caption{Example measurement using the polarized internal source after prepossessing using the CHARIS data-reduction pipeline. The red boxes indicate the apertures used to calculate the normalized Stokes parameters. The double-difference and double-sum images only have one side of the detector, since they are calculated by the difference and sum of the detector sides, respectively. When we align the detector sides, the apertures are effectively aligned as well.}
    \label{fig:aperture}
\end{figure}

The normalized Stokes parameters are calculated from the polarized internal source measurements as follows. For every observation and for every wavelength, the single sum and single difference are calculated by adding and subtracting, respectively, the images on the right and left sides of the detector (see \cref{eq:single_sum,eq:single_dif}). The single-sum and single-difference frames with the same derotator and HWP orientation are then combined by computing the pixel-to-pixel mean. We then compute frames of the double sum and double difference by combining the mean single sum and single difference frames with complementary single sum and single difference measurements (i.e., HWP difference of $45^\circ$ and the same derotator orientation, see \cref{eq:double_diff,eq:double_sum}). We then compute the trimmed mean, with a double-sided trimming fraction of 10\%, over rectangular apertures in the double-sum and double-difference frames. We place the apertures such that they do not include thermal effects at the edges of the detector and changes in the angular separation of orthogonal polarization states by the Wollaston prism as a function of wavelength. The apertures are shown in \cref{fig:aperture}. We use the trimmed mean to mitigate the effects of dead and hot pixels in the images. Outliers pull the mean from the center of the distribution, which results in an incorrect mean count value. Finally, we divide the double difference values by their corresponding double sum values to obtain the normalized Stokes parameters, according to equation \ref{eq:norm_stokes}.  

Using the Mueller matrix model introduced in \cref{cha:Model}, the optical path for the internal source measurements is described as:
\begin{equation}
    \boldsymbol{S}_\mathrm{det, L/R} = M_\mathrm{BS, L/R}T(-\Theta_\mathrm{der}) M_\mathrm{der} T(\Theta_\mathrm{der}) T(-\Theta_\mathrm{HWP}) M_\mathrm{HWP} T(\Theta_\mathrm{HWP}) T(\Theta_\mathrm{cal}) M_\mathrm{cal}\boldsymbol{S}_\mathrm{in},
\end{equation}
where $M_\mathrm{cal}$ is a Mueller matrix describing the calibration polarizer, characterized as $M_\mathrm{com}$ (\cref{eq:mueller_component}) and $\Theta_\mathrm{cal}$ is the rotation of the calibration polarizer including a misalignment angle, $\delta_\mathrm{cal}$, as $\Theta_\mathrm{cal} = \theta_\mathrm{cal} + \delta_\mathrm{cal}$. The diattenuation of the calibration polarizer is wavelength dependent. We assume the retardance of the calibration polarizer to be equal to $0^\circ$. The diattenuation of the calibration polarizer and the Wollaston prism are indistinguishable in our measurements. However, the extinction ratio of the Wollaston prism is very high (${>}$100000:1)\footnote{Thorlabs, Inc, \url{https://www.thorlabs.com/newgrouppage9.cfm?objectgroup_id=917}, consulted June 9, 2021.}. Therefore, we set the diattenuation of the Wollaston prism, $\epsilon_\mathrm{BS}$, equal to 1 in \cref{eq:mueller_calibrato1r} and we only fit the diattenuation of the calibration polarizer for all wavelengths separately. Using the polarized internal source measurements we can only measure the retardance of the derotator and HWP and not their diattenuation. To measure the diattenuation we would need to perform measurements with (nearly) unpolarized light. 
Therefore, the diattenuation of the derotator and HWP is set to 0 in our model. To fit the retardance of the derotator, one film width, $d$ is fitted for all three mirrors (see \cref{sec:derotator_model}). The three derotator mirrors rotate together, which makes their retarding effects indistinguishable. Thus, we can only estimate an effective film width for all three mirrors. To fit the retardance of the HWP, the width of the $\mathrm{SiO}_2$ crystal $w_{\mathrm{SiO}_2}$ and the width of the $\mathrm{MgF}_2$ crystal $w_{\mathrm{MgF}_2}$ are fitted separately (see \cref{sec:hwp_model}). We thus describe the instrumental polarization effects with three parameters for the retardance of the derotator and HWP, three parameters for the misalignment of the components ($\delta_\mathrm{cal}$, $\delta_\mathrm{der}$ and $\delta_\mathrm{HWP}$) and 22 parameters for the diattenuation of the polarizer ($\epsilon_{\mathrm{cal},\lambda}$), that is one for every wavelength bin.

We determine the model parameters by fitting the model function to the data points using non-linear least squares. The non-linear least squares equation is minimized using the Python implementation  \newline \verb|scipy.optimize.minimize|\cite{2020SciPy-NMeth}. We obtain the HWP and derotator angles from the rotation logs that are provided with the measurement. First, we perform fits where we directly fit the retardance of the HWP and derotator for all wavelengths separately, separately, which we refer to as the naive fits. These fits are used as initial guesses for the physical component model fits. 

\subsection{Results and Discussion of Internal Source Calibrations}\label{sec:internal_source_results}
The fitted model parameters and their $1\sigma$-uncertainties are shown in \cref{tab:instrument_model_parameters}. The uncertainties are calculated using the corrected sample standard deviation of the residuals, similar to the method used in Ref.~\citenum{vanHolstein2020IRDIS} (appendix E). For this calculation, the model parameters are assumed to be uncorrelated and do not have systematic errors. However, we see that the residuals are not normally distributed, which suggests that there are systematic errors in the model. Therefore, the error estimates should be interpreted as lower limits. 

\begin{table}[b]
    \centering
    \caption{Determined parameters and their errors of the components downstream of the telescope.}
    \label{tab:instrument_model_parameters}
    \begin{tabular}{||lc|c||}
        \hline
        \multicolumn{2}{||l|}{Parameter} & Value  \\ \hline
        $w_{\mathrm{SiO}_2}$  & $\mathrm{(mm)}$ & $1.623   \pm 0.001$  \\
        $w_{\mathrm{MgF}_2}$  & $\mathrm{(mm)}$ & $1.268   \pm 0.001$  \\
        $\delta_\mathrm{HWP}$ & $(^\circ)$      & $-0.002  \pm 0.004$  \\
        $d$     & $\mathrm{(nm)}$ & $262.56  \pm 0.03$   \\
        $\delta_\mathrm{der}$ & $(^\circ)$      & $-0.0118 \pm 0.0002$ \\
        $\delta_\mathrm{cal}$ & $(^\circ)$      & $  -0.035 \pm 0.009$ \\ \hline
    \end{tabular}
\end{table}

\begin{figure}
     \centering
     \begin{subfigure}[b]{0.49\textwidth}
         \centering
         \includegraphics[width=\textwidth]{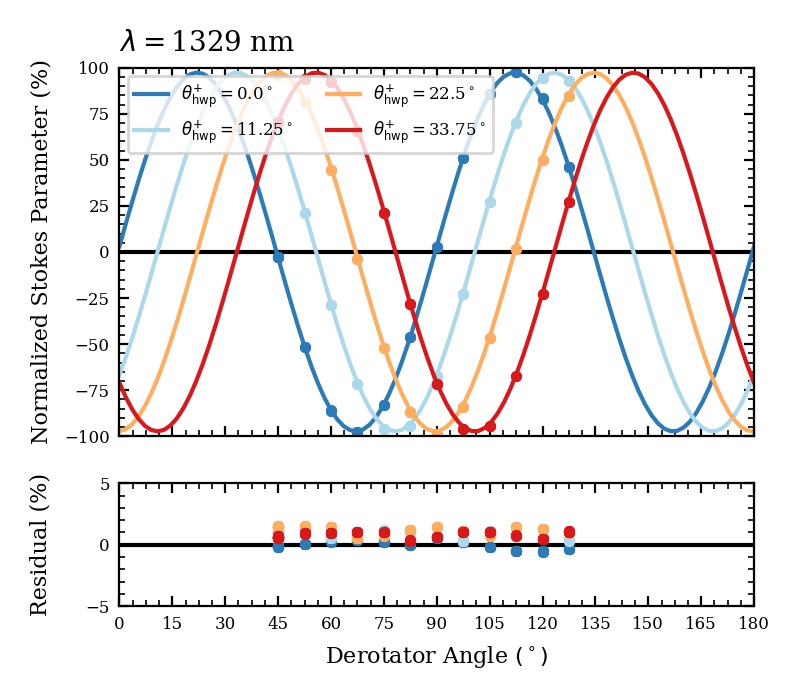}
     \end{subfigure}
     \hfill
     \begin{subfigure}[b]{0.49\textwidth}
         \centering
         \includegraphics[width=\textwidth]{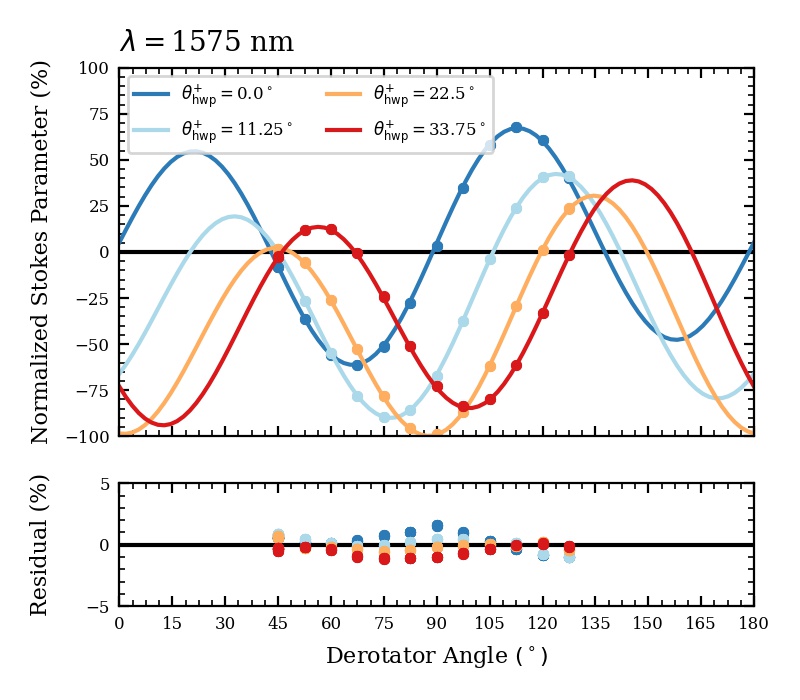}
     \end{subfigure}
     \hfill
     \begin{subfigure}[b]{0.49\textwidth}
         \centering
         \includegraphics[width=\textwidth]{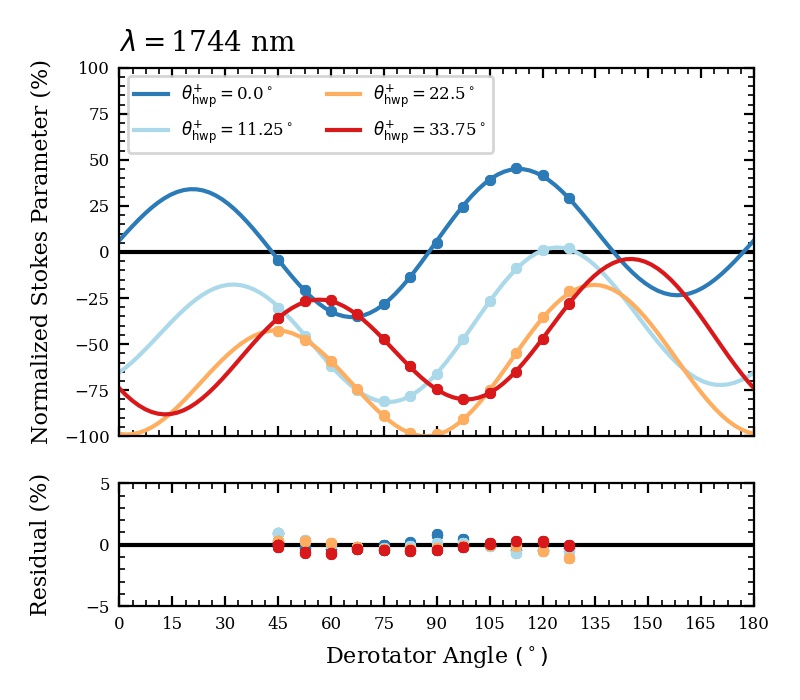}
     \end{subfigure}
     \begin{subfigure}[b]{0.49\textwidth}
         \centering
         \includegraphics[width=\textwidth]{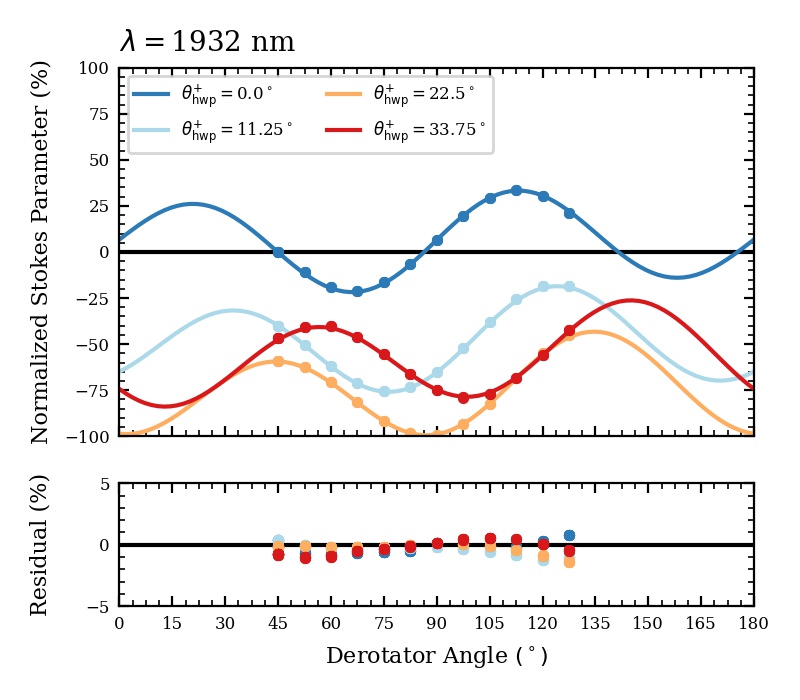}
     \end{subfigure}
        \caption{Measured and fitted normalized Stokes parameters as a function of derotator and HWP angle for four wavelength bins. The residuals of the fits are shown below the normalized Stokes parameters. The legend shows the $\theta_\mathrm{HWP}^+$ of the measurements, where it is implicit that $\theta_\mathrm{HWP}^-$ differs $45^\circ$.}
        \label{fig:internal_source_stokes_fits}
\end{figure}

\Cref{fig:internal_source_stokes_fits} shows the measured and fitted Stokes parameters for four wavelength bins. The legend of the figures indicates the $\theta_\mathrm{HWP}^+$ of the measurements. Recall that the normalized stokes parameters are calculated using two sets of observations with the HWP rotated $45^\circ$. Therefore, the corresponding $\theta_\mathrm{HWP}^-$ measurements differs $45^\circ$.  The residuals of the fits are shown below the normalized Stokes parameters. Overall, the model fits the data very well. However, the residuals of the fits show a systematic pattern that suggests a small offset between the data and the fit. The systematic errors could originate from errors in the observations or data reduction. Furthermore, the errors can be a result of the incompleteness of our model. The models for the diattenuation and retardance of the components can be oversimplified which makes it impossible for the model to exactly fit the data.

\begin{figure}
     \centering
     \begin{subfigure}[b]{0.49\textwidth}
        \centering
        \includegraphics[width=\textwidth]{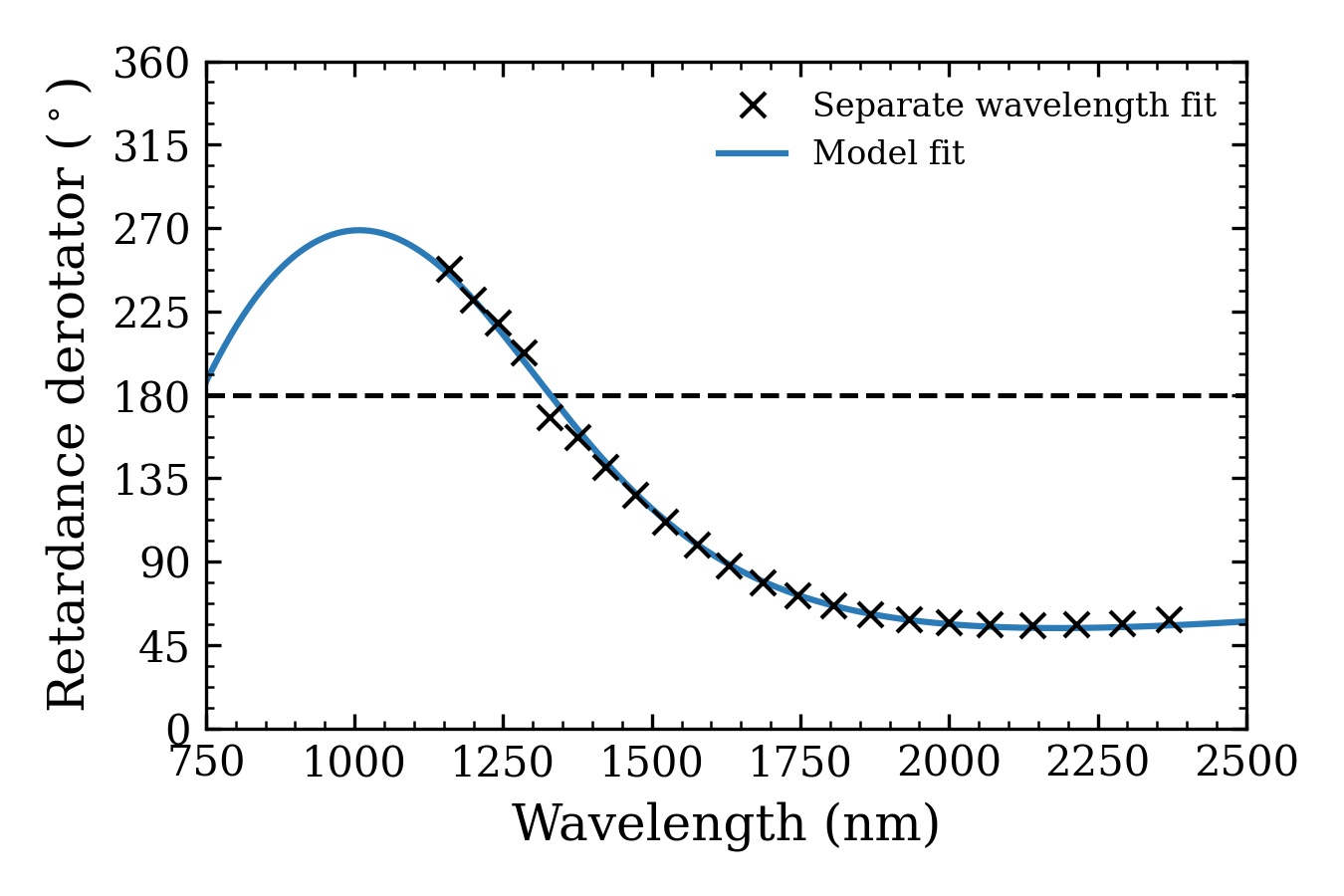}
        \caption{Derotator retardance model}
        \label{fig:retar_der}
    \end{subfigure}
    \hfill
    \begin{subfigure}[b]{0.49\textwidth}
        \centering
        \includegraphics[width=\textwidth]{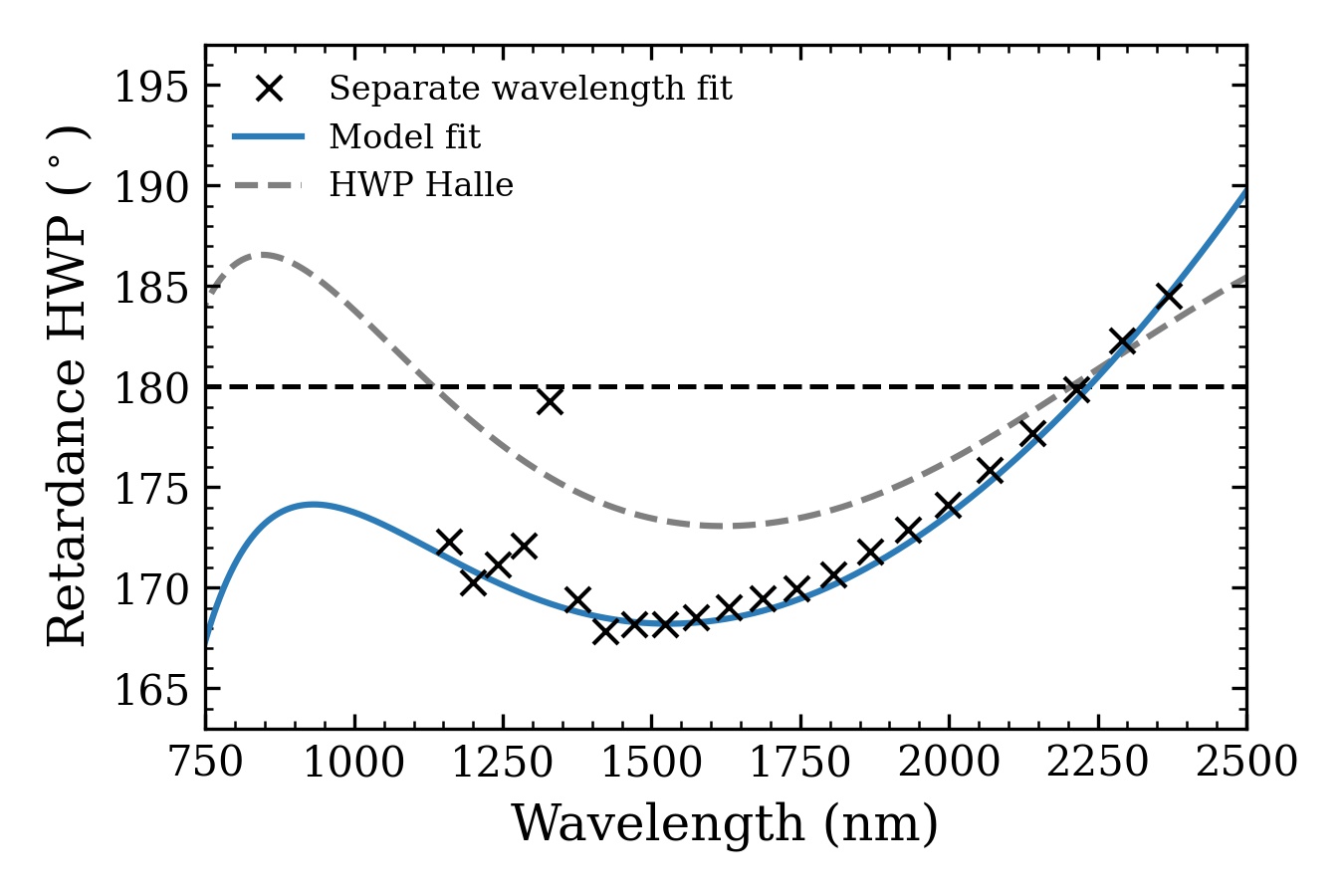}
        \caption{Half-wave plate retardance model}
        \label{fig:retar_hwp}
    \end{subfigure}
    \vspace{0.25cm}
    \caption{(a): Derotator retardance as a function of wavelength. Direct fits of the retardance for separate wavelengths are indicated by black crosses. The dark blue line indicates the model fit of the retardance, fitted using all wavelength bins simultaneously. With the black dashed line the ideal retardance of a perfect derotator is shown. (b) HWP retardance as a function of wavelength. The markers are the same as for (a). The gray dashed curve shows the retardance of the HWP used in SPHERE-IRDIS as given by the manufacturer.}
\end{figure}

\Cref{fig:retar_der,fig:retar_hwp} show the fitted models for the retardance of the derotator and HWP, respectively. In both figures, the results of the naive fits are indicated by the black crosses and the fit using the models defined in \cref{cha:Model} by the solid dark blue line. At $\lambda=1329\ \mathrm{nm}$, the naive fits of the retardance of the derotator and HWP differ more from the model fit than at other wavelengths. When the retardance of the derotator approaches the ideal value of $180^\circ$ a degeneracy arises in the retarding effects of the derotator and HWP. Therefore, the optimization cannot accurately fit both the retardance of the derotator and HWP at that wavelength. Breaking this degeneracy was one of the arguments to use the physical models for the derotator and HWP, besides reducing the number of model parameters. We do not know what specific HWP is used in SCExAO-CHARIS. However, to validate the HWP retardance model, the model is compared to the retardance of the HWP of SPHERE-IRDIS as specified by the manufacturer\footnote{B. Halle Nachfl. GmbH, \url{http://www.b-halle.de/products/retarders/achromatic_Retarders.html}, consulted June 9, 2021} (see \cref{fig:retar_hwp}). The curves are similar in shape. 

\begin{figure}
     \centering
     \begin{subfigure}[b]{0.49\textwidth}
         \centering
         \includegraphics[width=\textwidth]{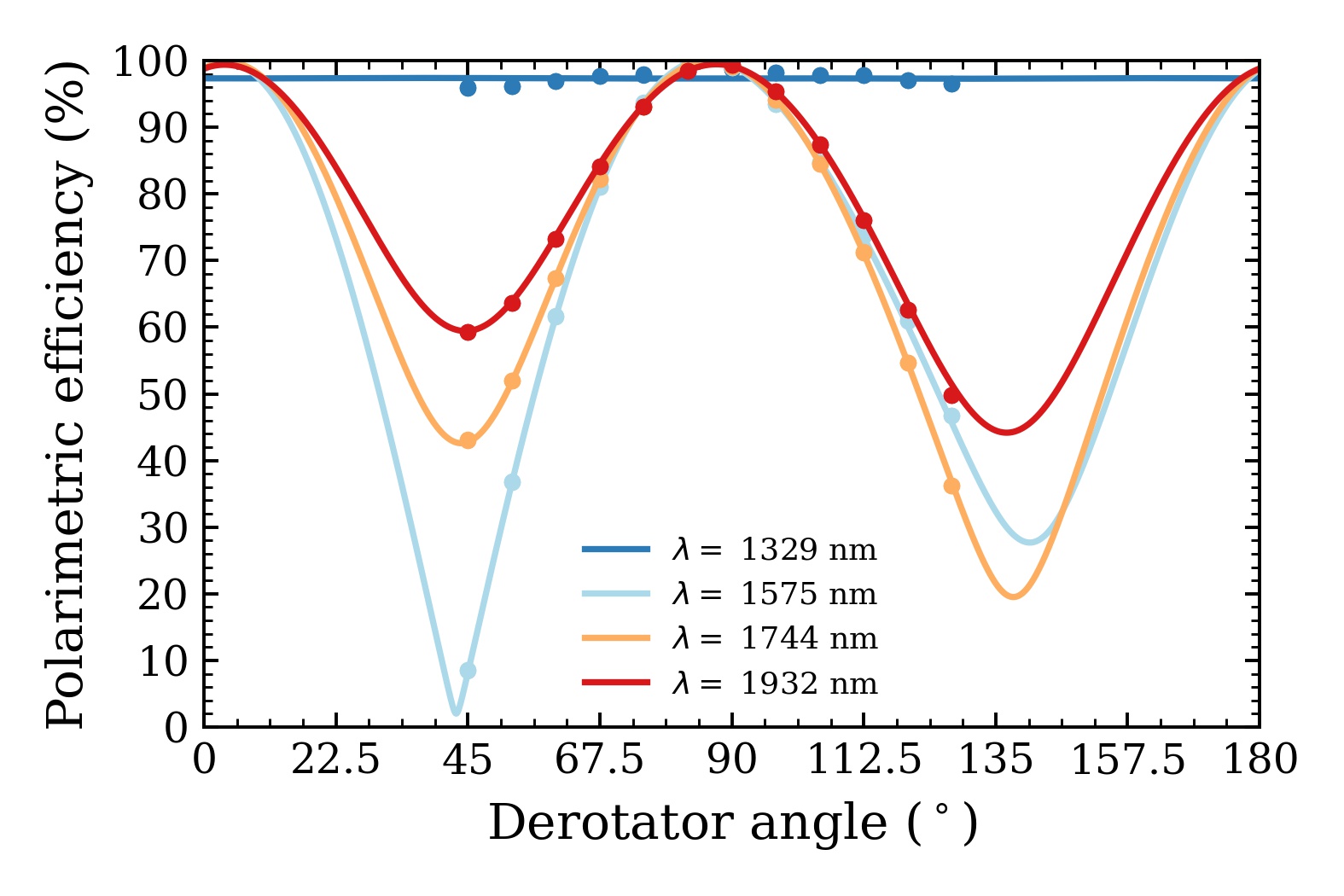}
        \caption{}
        \label{fig:dolp_internal_source}
     \end{subfigure}
     \hfill
     \begin{subfigure}[b]{0.49\textwidth}
         \centering
         \includegraphics[width=\textwidth]{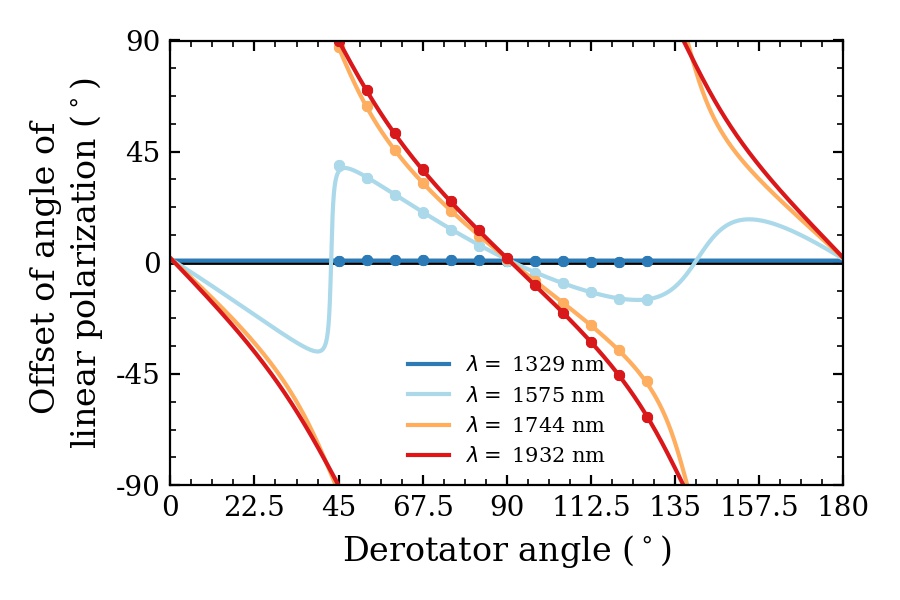}
        \caption{}
        \label{fig:aolp_internal_source}
     \end{subfigure}
     \vspace{0.25cm}
     \caption{Polarimetric efficiency (a) and offset of the angle of linear polarization (b) of the components downstream of the telescope as a function of wavelength for four wavelength bins. The normalized Stokes parameters, which are used to calculate (a) and (b), are calculated from measurements with $\theta^+_\mathrm{HWP} = (0^\circ, 22.5^\circ)$.}
\end{figure}

\begin{figure}[h]
     \centering
     \begin{subfigure}[t]{0.49\textwidth}
        \centering
        \includegraphics[width=\textwidth]{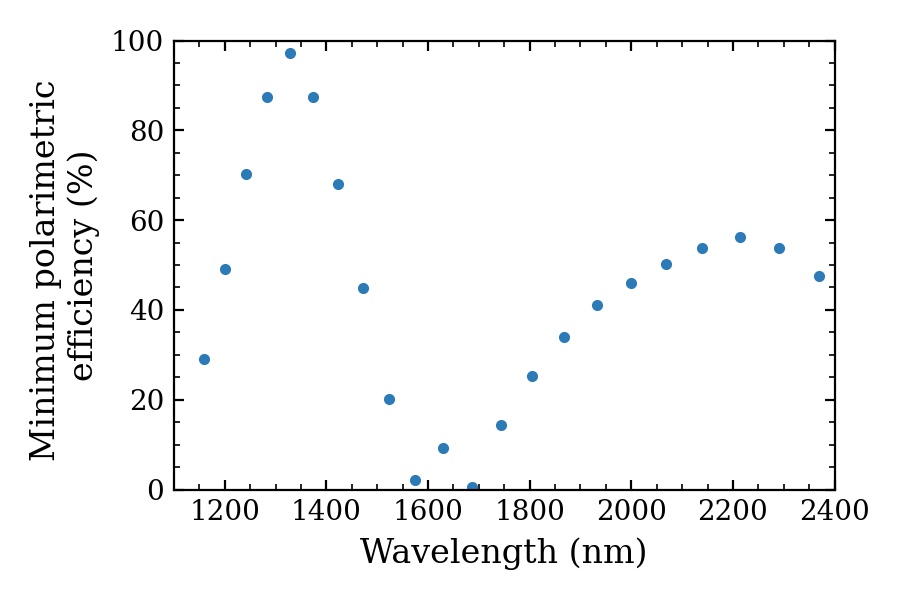}
    \caption{}
    \label{fig:min_dolp}
    \end{subfigure}
    \hfill
    \begin{subfigure}[t]{0.49\textwidth}
        \centering
        \includegraphics[width=\textwidth]{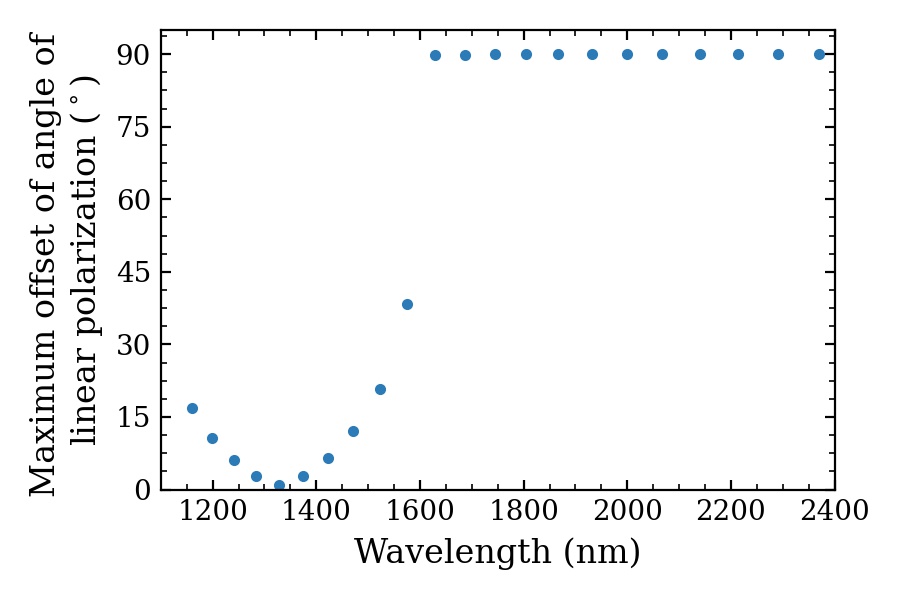}
        \caption{}
        \label{fig:max_aolp}
    \end{subfigure}
    \vspace{0.25cm}
    \caption{Minimum polarimetric efficiency (a) and maximum offset of the angle of linear polarization (b) as a function of wavelength.}
\end{figure}

To illustrate the effect of the determined instrumental polarization effects, \cref{fig:dolp_internal_source} shows the measured and fitted degree of linear polarization $P$ as a function of derotator angle for four wavelength bins. Since the light generated by the calibration polarizer is nearly 100\% polarized, we can interpret $P$ as the polarimetric efficiency, that is, the fraction of true linear polarization that is actually measured at the detector. Recall that the data points in \cref{fig:internal_source_stokes_fits} are normalized Stokes parameters calculated using the double-sum and double-difference using pairs of observations with $\theta_\mathrm{HWP}$ differences of $45^\circ$. we calculate $P$ from \cref{eq:dolp} using pairs of Stokes parameters for which their $\theta_\mathrm{HWP}^+$ differ $22.5^\circ$ or $67.5^\circ$. The curves and data points shown in \cref{fig:dolp_internal_source} are computed with $\theta_\mathrm{HWP}^+ = 0^\circ$ and $\theta_\mathrm{HWP}^+ = 22.5^\circ$ (and correspondingly $\theta_\mathrm{HWP}^- = 45^\circ$ and $\theta_\mathrm{HWP}^- = 67.5^\circ$, respectively). 

The polarimetric efficiency of the $\lambda = 1329$ nm bin is not a perfect fit. The measured polarimetric efficiency shows a curve pattern, similar to the other wavelength bins, but the model fits an almost straight line. The model of the derotator retardance predicts that the retardance of the derotator at wavelengths close to $1329$ nm is very close to $180^\circ$. Therefore, the model predicts that the polarimetric efficiency of the system should be nearly perfect, for all derotator orientations, resulting in a straight line. The offset of the line is a result of the fit of the diattenuation of the polarizer. 

\Cref{fig:dolp_internal_source} shows that at $\lambda = 1575\ \mathrm{nm}$ the polarimetric efficiency dramatically drops for derotator angles close to $45^\circ$ and $135^\circ$. The low efficiency is due to the derotator retardance strongly deviating from the ideal value of $180^\circ$. \Cref{fig:min_dolp} shows the minimum polarimetric efficiency as a function of wavelength. We calculate the minimum polarimetric efficiency by calculating the minimum of model fits. At wavelengths close to $1600\ \mathrm{nm}$ the derotator retardance is close to $90^\circ$, which makes it act similar to a quarter-wave plate. Therefore, at 1600~$\mathrm{nm}$ and $\theta_\mathrm{der} \approx 45^\circ$ (and $\theta_\mathrm{der} \approx 135^\circ$) almost all incident linearly polarized light is converted to circularly polarized light, resulting in a very low polarimetric efficiency of only several percent. This behavior is similar to the behavior seen at SPHERE-IRDIS\cite{vanHolstein2020IRDIS}. The strong wavelength dependence in \cref{fig:min_dolp} is due to the wavelength dependence in the retardance of the derotator. Derotator rotation angles of $45^\circ$ and $135^\circ$ therefore need to be avoided during observations. 

The effect of the retardance of the HWP is much smaller since the retardance is for the entire wavelength range relatively close to the ideal $180^\circ$. The effect of the HWP's retardance is visible in \cref{fig:dolp_internal_source} as the variation in the skewness of the curves and the asymmetry of the heights and shapes of the curves at $\theta_\mathrm{der} \approx 45^\circ$ and $\theta_\mathrm{der} \approx 135^\circ$. The horizontal shifting of the curves is caused by the misalignment of the calibration polarizer, HWP and derotator ($\delta_\mathrm{cal}$, $\delta_\mathrm{HWP}$ and $\delta_\mathrm{der}$). 


Apart from reducing the polarimetric efficiency, the crosstalk produced by the derotator and HWP also causes an offset in the angle of linear polarization, $\chi$, with respect to the perfect case. \Cref{fig:aolp_internal_source} shows the measured and fitted offset in $\chi$ for four wavelength bins. The offsets are calculated as the measured/fitted $\chi$ minus the expected $\chi$ for a perfect HWP and derotator. \Cref{fig:max_aolp} shows the maximum offset of $\chi$ as a function of wavelength. We calculated the maximum offset of $\chi$ by calculating the maximum of the model fits. For $\lambda \gtrsim 1600\ \mathrm{nm}$ the maximum offset is $90^\circ$. The non-ideal HWP retardance causes the asymmetry at $\theta_\mathrm{der} \approx 45^\circ$ and $\theta_\mathrm{der} \approx 135^\circ$.

\begin{figure}
    \centering
    \includegraphics[width=0.49\textwidth]{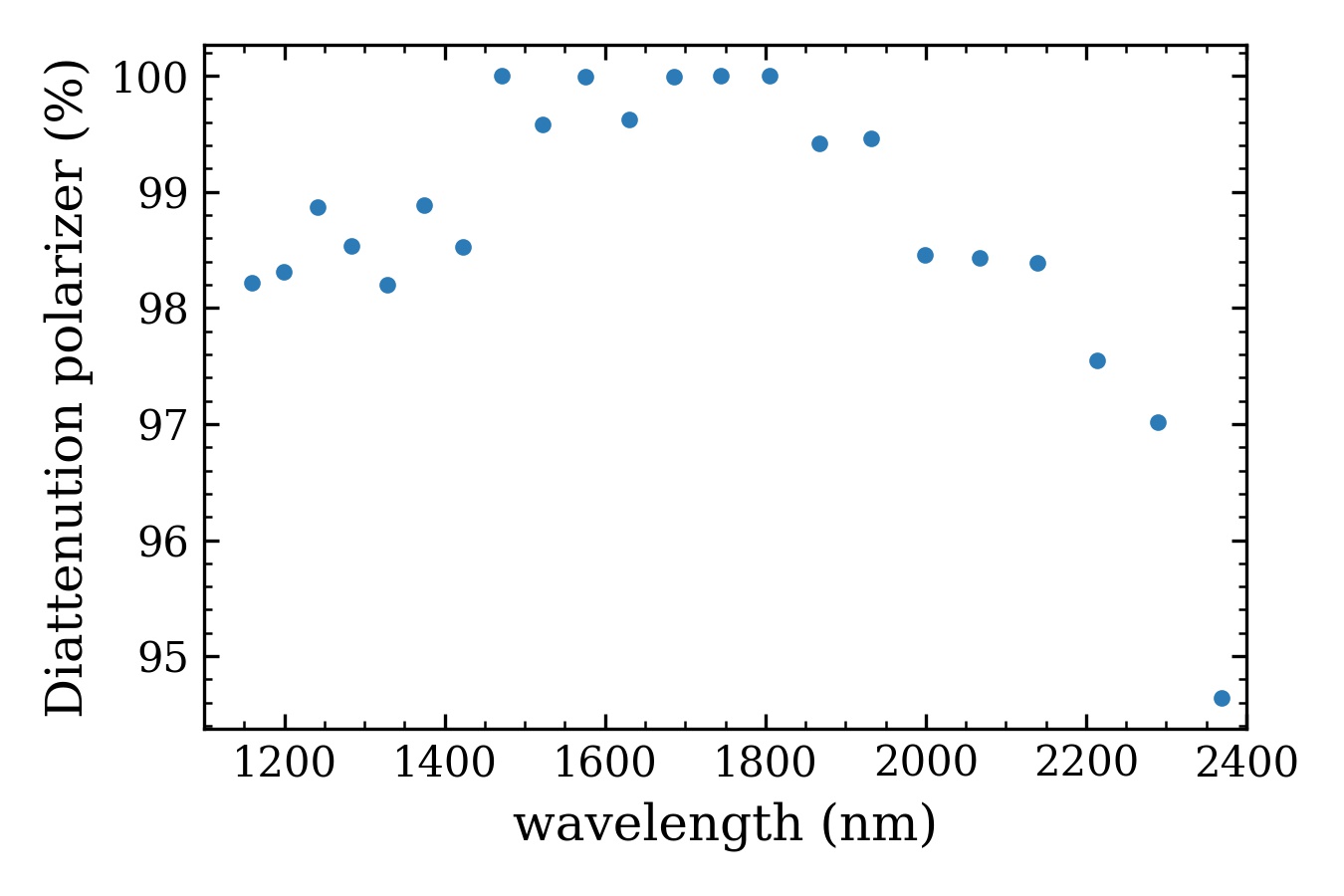}
    \caption{Fitted diattenuation of the calibration polarizer as a function of wavelength. The uncertainty on the fitted diattenuation is ${\sim} 0.2\%$.}
    \label{fig:diatt_cal}
\end{figure}

The fitted diattenuation of the calibration polarizer is shown in \cref{fig:diatt_cal}. The diattenuation of the calibration polarizer does not show a strong wavelength dependence. We argue that this is an artifact of the model and fitting method. Since no wavelength-dependent constraints are imposed on the diattenuation of the calibration polarizer, these parameters are the least constrained of all model parameters. Therefore, the model tries to correct for imperfections in the data or in the model by changing the diattenuation of the calibration polarizer. A mathematical description of the wavelength dependence of the calibration polarizer proved non-trivial and is left out of the scope of this project. The imperfect fit of the diattenuation of the calibration polarizer will have minor effects on the final correction model.

Compared to Ref.~\citenum{vanHolstein2020IRDIS}, the fitted $P$ and offset of $\chi$ are less accurate for the individual wavelength bins. However, using the physical models we mitigate systematic errors in individual wavelength bins. Furthermore, the physical models strongly reduce the number of model parameters. In Ref.~\citenum{vanHolstein2020IRDIS}, the retardance and diattenuation of the derotator and HWP of SPHERE-IRDIS are fitted directly for all wavelength bins separately, which makes the model more susceptible for systematic errors in the calibration data. However, our model has fewer degrees of freedom, which makes it more difficult to accurately fit the data. 

The IP of the HWP and derotator are expected to be low. The calibration measurements of SPHERE-IRIDS showed that the diattenuation of the HWP is ${<}0.1\%$ and the diattenuation of the derotator ${<}1\%$ for all wavelengths\cite{vanHolstein2020IRDIS}. The retardance fits of the HWP and derotator in this work are similar to the values obtained for SPHERE-IRIDS. Therefore, we expect the diattenuation to be similar as well. Since we have used a physical model based on the Fresnel equations for the derotator, we can estimate the diattenuation of the derotator using the model parameters obtained with the internal source measurements. The estimated diattenuation of the derotator is shown in \cref{cha:appendix_der_diatt}.


%% file: onsky.tex
Now that we have a model describing the optical components downstream of the telescope, we can investigate the instrumental polarization effects of the telescope mirrors. To this end, we analyze observations of an unpolarized standard star. In \cref{sec:on_sky_methods} we discuss the data and the methods used to estimate the model parameters and in \cref{sec:on_sky_results} we present and discuss the results of the fits of the model parameters.

\subsection{Calibration Measurements and Estimation of Model Parameters}\label{sec:on_sky_methods}
To determine the IP of the telescope mirrors, we observed the unpolarized star HD~140667 during a SCExAO engineering night. HD~140667 has a measured degree of linear polarization ($P$) of $28 \pm 8\ \mathrm{ppm}$ in V-band\cite{piirola2020standardpol} and is at a distance of 41 pc\cite{skiff2014vizier}. 
On the night of March 21, 2021, two sets of observations were taken at different altitude angles. The first set of measurements are performed at altitude angles ranging from $46^\circ$ to $50^\circ$ and the second set at altitude angles ranging from $77^\circ$ to $81^\circ$. The degree of polarization, $P$, is expected to be constant for all altitude angles because the telescope mirrors rotate as a whole. 
However, as seen from the Nasmyth platform the telescope pupil image is rotated with minus the altitude angle, thus changing the amount of $Q$ and $U$ as seen from SCExAO-CHARIS. By measuring $Q$ and $U$ over a broad range of altitude angles we can constrain the model parameters to high accuracy. 

During the observations, we fixed the derotator at $90^\circ$ for high polarimetric efficiency. We used the adaptive optics (AO) in a mode that gives only a marginal wavefront correction to reach a high total photon count. 
This way we could integrate longer without saturating or reaching the non-linearity regime of the detector. The integration time of each frame is ${\sim} 4$ seconds. We moved the coronagraphs out of the beam and we used no ND filters. 
We rotated the HWP relative to the fixed instrument reference frame.
We obtained six HWP cycles ($\theta_\mathrm{HWP}=0^\circ$ and $45^\circ$ for measuring Stokes $Q$ and $\theta_\mathrm{HWP}=22.5^\circ$ and $67.5^\circ$ for measuring Stokes $U$) of which three at low altitude angles and three at high altitude angles. We used the CHARIS data-reduction pipeline with the standard settings to preprocess the measurements\cite{brandt2017data}.



The seeing conditions were variable at the night of the observations and a broad speckle pattern is visible in many of the frames. Many frames are saturated because the conditions improved over time. According to Ref.~\citenum{brandt2017data}, the detector response becomes non-linear at ${\sim}40,000$ counts. The CHARIS data reduction pipeline corrects for non-linearity of the detector, but we decided to mask all pixels with counts higher than $40,000$ counts to prevent saturation and non-linearity effects contaminating the data. The mask only masks the center of the PSF core and masks a maximum of 36 pixels.

Due to artifacts in the CHARIS data reduction pipeline, varying noise is visible at the bottom of the frames. The noise is far enough from the source to not contaminate the observed flux of the star. However, we masked the bottom of all frames when calculating the background of the frames. 

We center the frames using a Gaussian fit, which is a sufficiently good approximation of a seeing-limited PSF. From the centered frames, we calculate the single difference and single sum using \cref{eq:single_dif,eq:single_sum}, respectively. We then combine the single-difference and single-sum frames in the same HWP cycle and with the same HWP orientation by taking the mean of the frames. Using the mean-single-sum and mean-single-difference frames, we subsequently calculate the double-difference and double-sum images using equation \cref{eq:double_diff,eq:double_sum}, respectively. We mask part of the double-difference and double-sum frames to mitigate saturation as follows. We calculate the masks for all frames and wavelength bins separately. We then combine all masks of the frames used to calculate the double-difference and double-sum images by masking all pixels that are masked at least once in any of the masks. We then apply these masks to the double-difference and double-sum frames and calculate the flux in these frames in an aperture with a radius of 25 pixels. 

\begin{table}[b]
    \centering
    \caption{Best fit parameters for the M3 model. The parameters describe the parameterization of the approximation of the dielectric functions.}
    \label{tab:telescope_model_parameters}
    \begin{tabular}{||l|l||}
        \hline
        Parameter               & Value  \\ \hline
        $m_\mathrm{1}$     & 2.104  \\
        $b_\mathrm{1}$     & 14.20  \\
        $m_\mathrm{2}$     & 2.100  \\
        $b_\mathrm{2}$     & 13.20  \\ \hline
    \end{tabular}
\end{table}

We calculate the normalized Stokes parameters from the observations as follows. In order to estimate the background in the double sum and double difference, we look at the radial profile of the flux in the double-sum and double-difference images. In the ideal case, the background would reach a constant value at a radius far from the source. The double-difference counts in an annulus go to zero for radii far from the center of the source. However, the counts of the double-sum images in an annulus at large separation do not go to a constant value but keep decreasing, because the PSF is larger than the FOV of the detector. This is also observed by Ref.~\citenum{vanHolstein2020IRDIS} (appendix D). Similarly to Ref.~\citenum{vanHolstein2020IRDIS}, we calculate the background in both the double-difference and the double-sum images inside an annulus as far from the source as possible. In our case, this is an annulus with an inner radius of 55 pixels and an outer radius of 60 pixels. Effectively, this annulus is an annulus sector that only covers the top part of the detector, because the bottom part is masked and the annulus falls outside the image to the left and right of the image. The normalized Stokes parameters are calculated using \cref{eq:norm_stokes}. This method introduces a slight overestimation of the background flux in total intensity, thus lowering the total double-sum count. The effect is however small due to the high double-sum count in all frames.

Using the compiled data set, we estimate the diattenuation of the telescope M3 mirror by fitting a linear approximation of the dielectric functions (see \cref{sec:m3_model}). Recall that the diattenuation of M1 and M2 are assumed to be minimal. Similar to the internal source calibration, we first fit the diattenuation of M3 for all wavelengths separately. We again call these fits naive fits and use them as initial guesses for the physical model. We perform the fits as discussed in \cref{sec:internal_source_methods}.

\begin{figure}[b]
     \centering
     \begin{subfigure}[b]{0.49\textwidth}
         \centering
         \includegraphics[width=\textwidth]{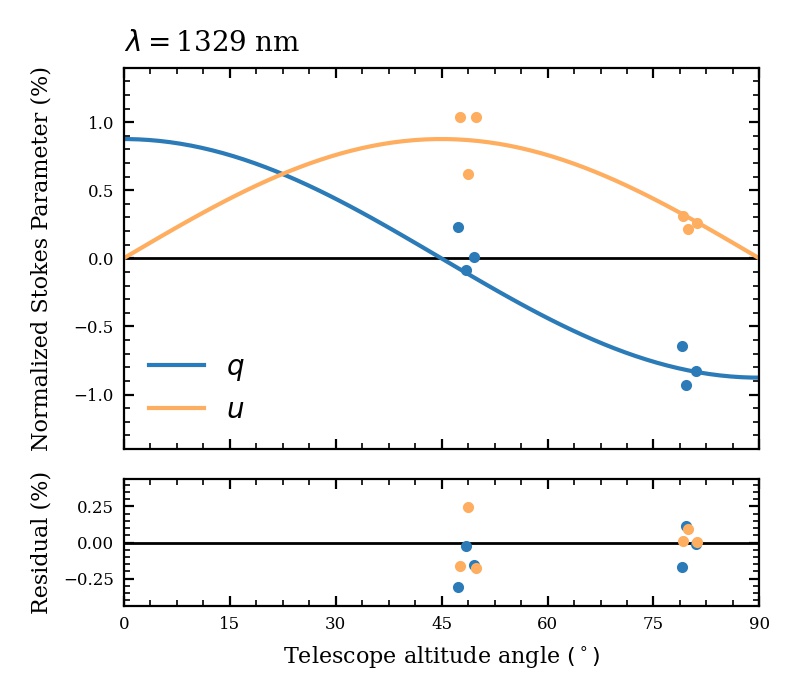}
     \end{subfigure}
     \hfill
     \begin{subfigure}[b]{0.49\textwidth}
         \centering
         \includegraphics[width=\textwidth]{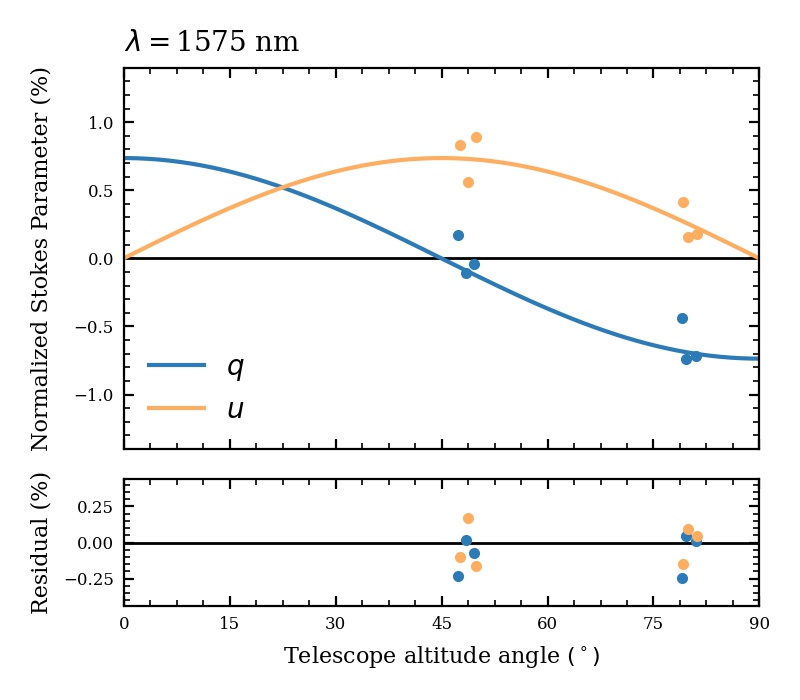}
     \end{subfigure}
     \hfill
     \begin{subfigure}[b]{0.49\textwidth}
         \centering
         \includegraphics[width=\textwidth]{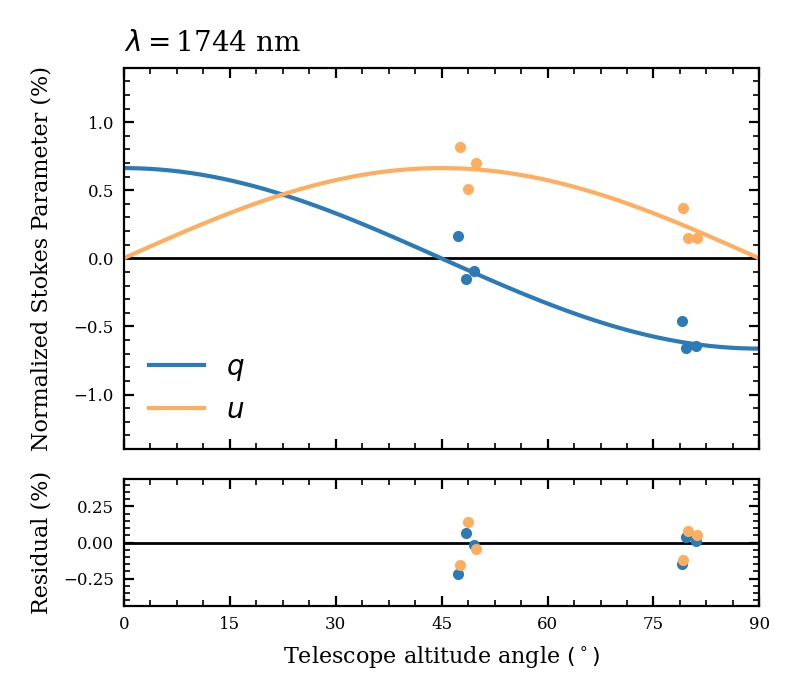}
     \end{subfigure}
     \begin{subfigure}[b]{0.49\textwidth}
         \centering
         \includegraphics[width=\textwidth]{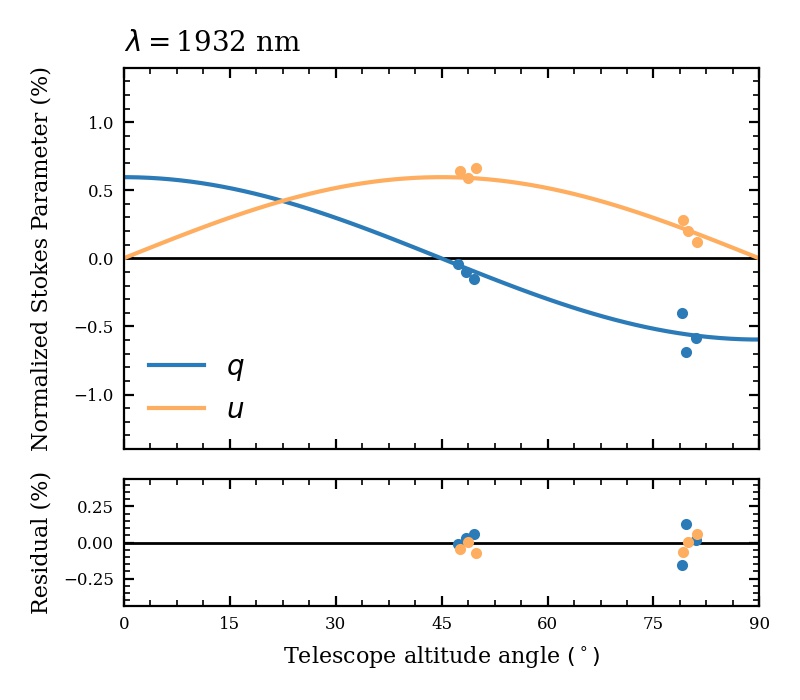}
     \end{subfigure}
        \caption{Measured and fitted normalized Stokes parameters as a function of altitude angle for four wavelength bins. Below the normalized Stokes parameters, the residuals of fit are shown for the four wavelength bins.}
        \label{fig:on_sky_stokes_fits}
\end{figure}

\subsection{Results and Discussion of Unpolarized Star Calibration}\label{sec:on_sky_results}
The fitted model parameters are presented in  \cref{tab:telescope_model_parameters}. We attempted the same approach as for the internal source measurements to estimate the errors on the parameters. However, the estimate of the Jacobian matrix proved to be near singular. We have therefore not been able to estimate the errors on the fitted parameters. Furthermore, the thermal noise on the longest wavelength bin, $\lambda = 2369\ \mathrm{nm}$, is such that we are unable to perform a good fit to that data. Therefore, we exclude this wavelength bin from all analyses of the IP of the telescope mirrors.

\begin{figure}
    \centering
    \includegraphics[width=0.49\textwidth]{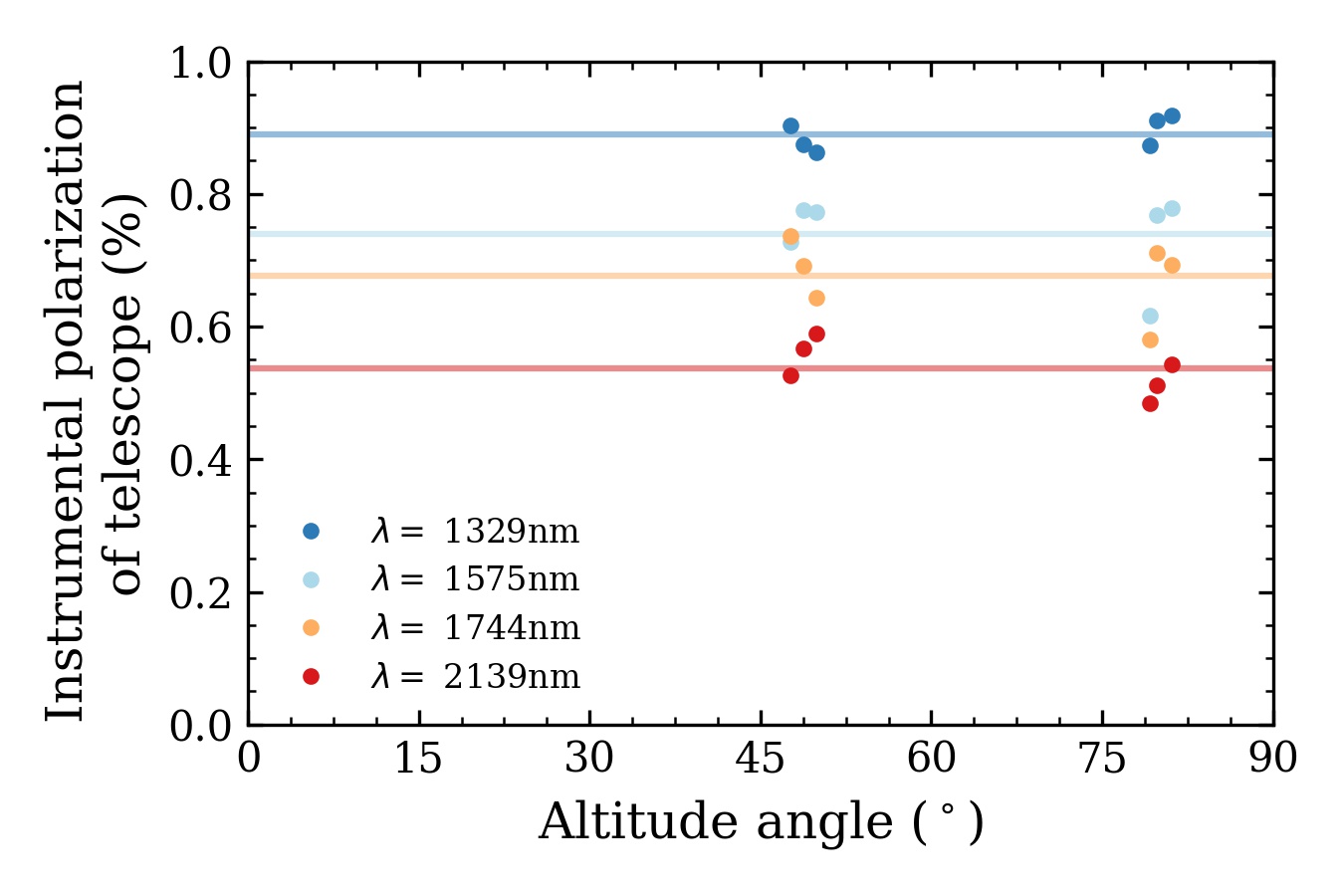}
    \caption{Degree of linear polarization $P$ a function of altitude angle for four wavelength bins. The horizontal lines show the mean $P$ of the different wavelengths.}
    \label{fig:dolp_onsky_alt}
\end{figure}

\Cref{fig:on_sky_stokes_fits} shows the normalized Stokes parameters and the model fit as a function of altitude angle for four wavelength bins. The scatter in the data points is larger than for the internal source measurements. This is primarily caused by the varying seeing during the observations. There are also systematic errors visible in the data points. For instance, at low altitude, the $u$ data points always show a `triangular'-like structure. We have tried to use the normalized double-difference to reduce the noise in the normalized Stokes parameters, as suggested in Ref.~\citenum{vanHolstein2020IRDIS}. The normalized double difference reduced the systematic errors visible in \cref{fig:on_sky_stokes_fits}. However, it introduced new systematic errors. The model seems to underestimates the Stokes parameters calculated with the normalized double difference. Furthermore, the $q$ data points at low altitude angles had a systematic offset from zero, which is not expected. In the end, we decided to use the standard double difference to calculate the normalized Stokes parameters.

\Cref{fig:dolp_onsky_alt} shows the degree of linear polarization, $P$, as a function of telescope altitude angle. The $P$ of the observations, and hence the IP of the telescope, do not show a dependence on the altitude angle. With the horizontal lines, we show the mean $P$ for the different wavelengths. 
This constant IP is different to that of SPHERE-IRDIS, where the IP increases for lower altitude angles \cite{vanHolstein2020IRDIS}. This is because the IP of SPHERE-IRDIS results from the M3 of the telescope, which rotates with the altitude angle with respect to the instrument, as well as a fixed fold mirror within the instrument. For SCExAO-CHARIS, the IP is the result of M3 only, and therefore the IP is constant with altitude angle.

\begin{figure}
    \begin{subfigure}[b]{0.49\textwidth}
        \centering
        \includegraphics[width=\textwidth]{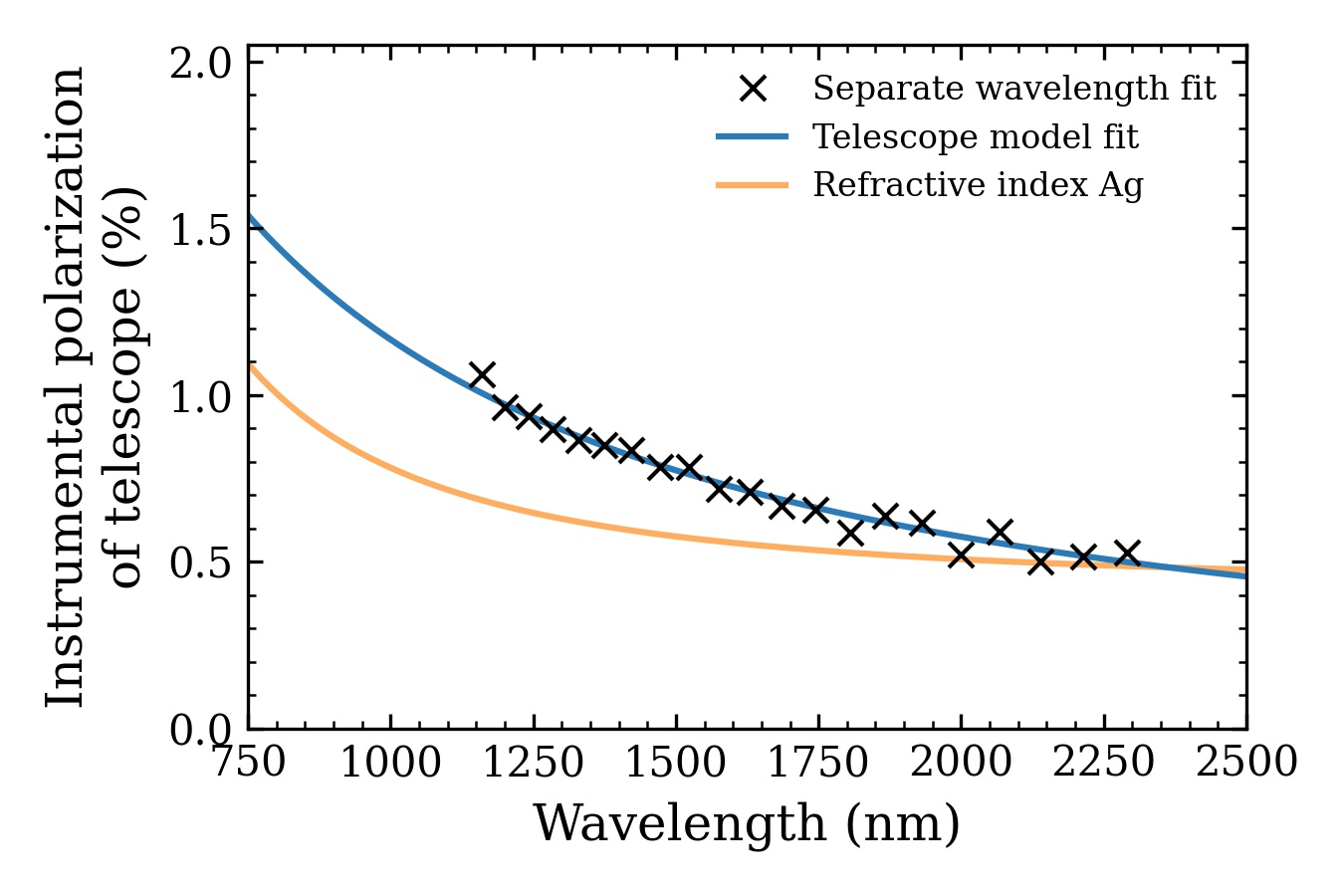}
        \caption{}
        \label{fig:diatt_tel}
    \end{subfigure}
    \hfill
    \begin{subfigure}[b]{0.49\textwidth}
        \centering
        \includegraphics[width=\textwidth]{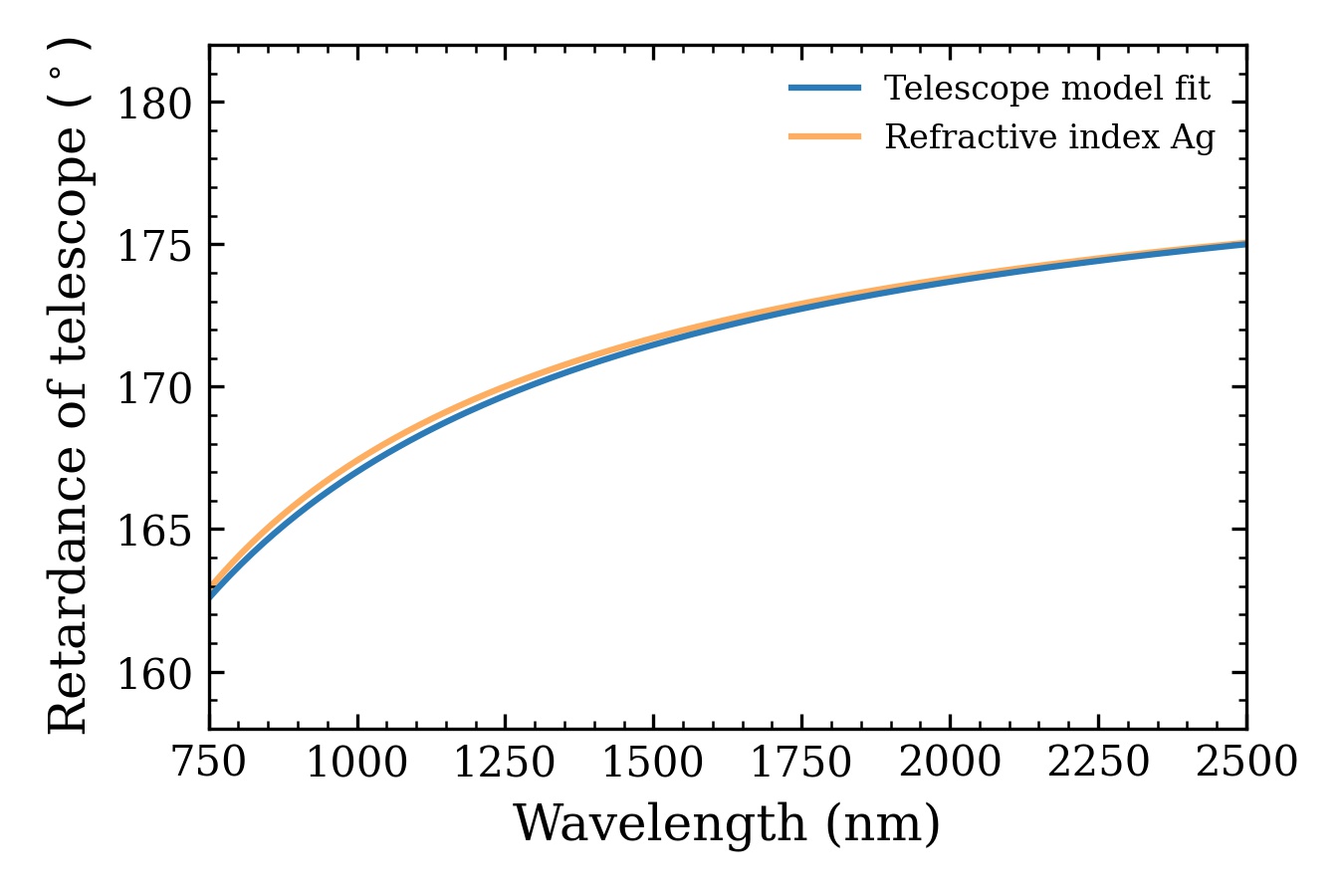}
        \caption{}
        \label{fig:retar_tel}
    \end{subfigure}
    \vspace{0.25cm}
    \caption{The fitted diattenuation (a) and derived retardance (b) of the telescope using the M3 model. In (a), the black crosses indicate the direct diattenuation fits for all wavelength bins, separately, The dark blue curve show the model fit and the orange line the expected diattenuation of a perfect silver mirror. Note that the diattenuation can be directly interpreted as the IP of the telescope mirrors. (b) shows the derived retardance from the fitted effective refractive index and the retardance of a perfect silver mirror. As the retardance of the telescope cannot be measured from observations of an unpolarized source, no direct fits are shown.}
\end{figure}

\Cref{fig:diatt_tel} shows the fitted diattenuation of the M3 mirror. The diattenuation of M3 can be interpreted as the IP generated by the telescope. The black crosses show the naive fits and the solid dark blue line shows the model fit. The naive fits are noisier at longer wavelengths, which is due to an increase in thermal background noise for these wavelengths. The orange line shows the expected diattenuation for a bare silver mirror at $45^\circ$ incidence\cite{rakic1998optical}. The fitted diattenuation of the telescope is higher than the theoretically predicted diattenuation, causing more IP. The fitted and theoretical refractive indices are very similar. This shows that small differences in the refractive index result in significant differences in IP. Therefore, it is clear that calibration measurements are always necessary to calibrate for the IP of telescope mirrors, instead of using models based on literature values\cite{witzel2011calibration}.

We cannot measure the retardance of the telescope mirrors using measurements of an unpolarized target. To measure the retardance of the telescope, observations from a polarized source are needed. Since our model of the telescope mirrors fit the effective refractive index of the telescope mirror we can make an estimate of the retardance of the telescope mirrors. The retardance is expected to be almost completely determined by reflection from M3 (see \cref{sec:polarization_effects_of_instrument}). \cref{fig:retar_tel} shows the retardance of a mirror based on the fitted refractive index and with the theoretical refractive index of bare silver of a mirror at $45^\circ$ incidence. The theoretical and fitted curves are very close together. The retardance of a mirror predominantly depends on the imaginary part of the refractive index $\kappa$. Since the theoretical and fitted $\kappa$ are very similar, the theoretical and fitted retardance is very similar. Even though we did not measure the retardance directly, the derived retardance is most likely more accurate than the retardance calculated for the telescope mirrors of SPHERE-IRDIS\cite{vanHolstein2020IRDIS}. This is because we have estimated the retardance of M3 using observations, whereas for the Mueller matrix model of SPHERE-IRDIS literature values have been used. 


%% file: accuracy.tex
\begin{figure}
    \centering
    \begin{subfigure}[b]{0.49\textwidth}
        \centering
        \includegraphics[width=\textwidth]{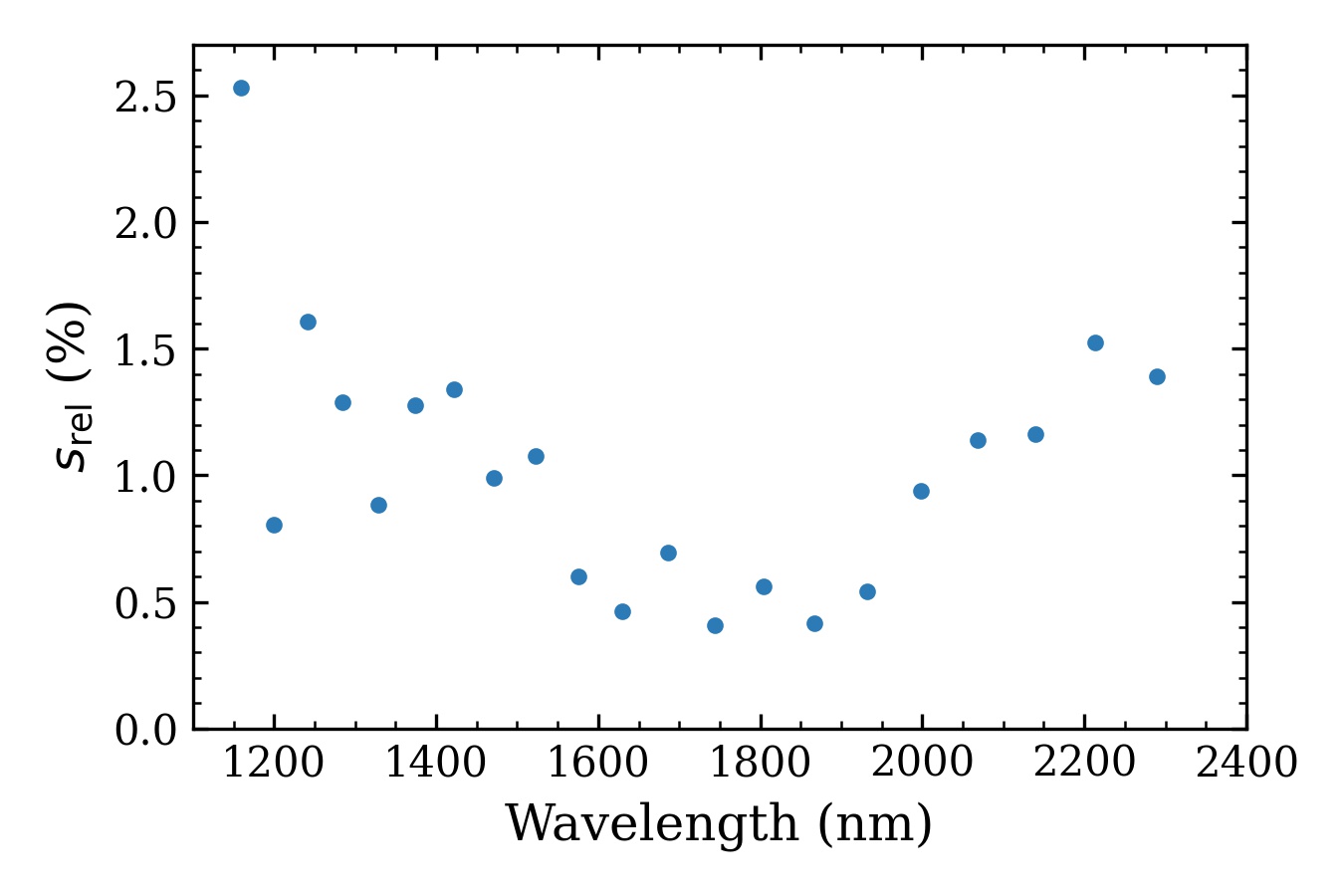}
        \caption{}
        \label{fig:s_rel}
    \end{subfigure}
    \hfill
    \begin{subfigure}[b]{0.49\textwidth}
        \centering
        \includegraphics[width=\textwidth]{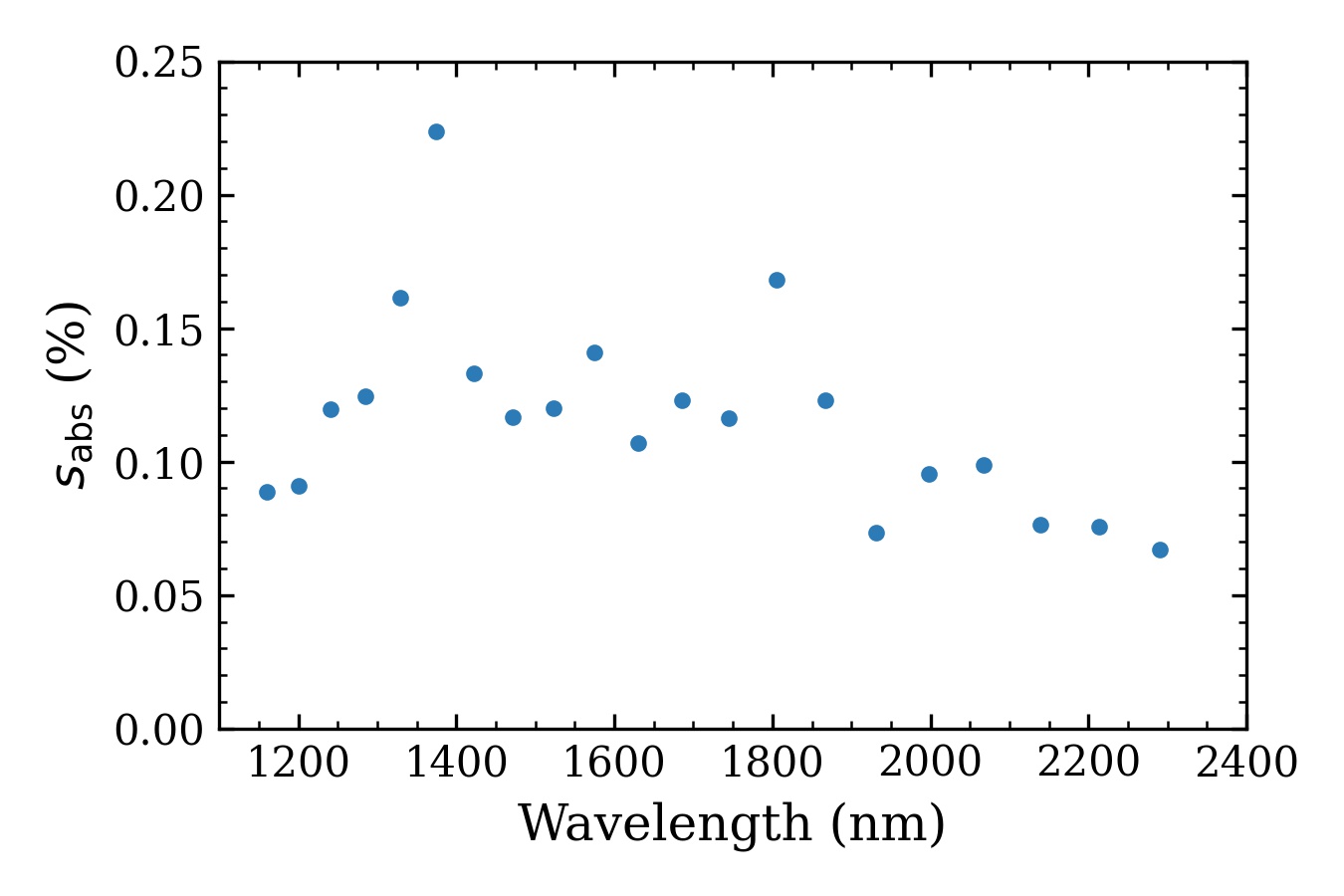}
        \caption{}
        \label{fig:s_abs}
    \end{subfigure}
    \vspace{0.25cm}
    \caption{Relative (a) and absolute (b) polarimetric accuracies as a function of wavelength. The relative accuracy is determined from the measurements using the internal source and  the absolute accuracy is determined from the measurements of the unpolarized star.}
\end{figure}

Now that we have a total description of the instrumental polarization effects of the entire instrument we will look at the accuracy of our model. This method is based on the method described in Ref.~\citenum{vanHolstein2020IRDIS} (appendix E). We start with estimating the accuracy of the fit of the model parameters to the calibration data using the corrected sample standard deviation of the residuals. To compute the total polarimetric accuracy, we compute the absolute polarimetric accuracy, $s_\mathrm{abs}$, that is the uncertainty in the IP and the relative polarimetric accuracy, $s_\mathrm{rel}$, that is the uncertainty that scales with the input signal. We use the accuracy of the fit on the calibration data of the unpolarized star as $s_\mathrm{abs}$ and the accuracy of the fit on the calibration data with the polarized internal source as $s_\mathrm{rel}$. In \cref{fig:s_rel,fig:s_abs}, $s_\mathrm{rel}$ and $s_\mathrm{abs}$ are shown as a function of wavelength. Both $s_\mathrm{rel}$ and $s_\mathrm{abs}$ show a quite strong wavelength dependence and range from $0.41\%$ to $2.5\%$ and from $0.067\%$ to $0.22\%$, respectively. 

\begin{figure}
     \centering
     \begin{subfigure}[b]{0.49\textwidth}
         \centering
         \includegraphics[width=\textwidth]{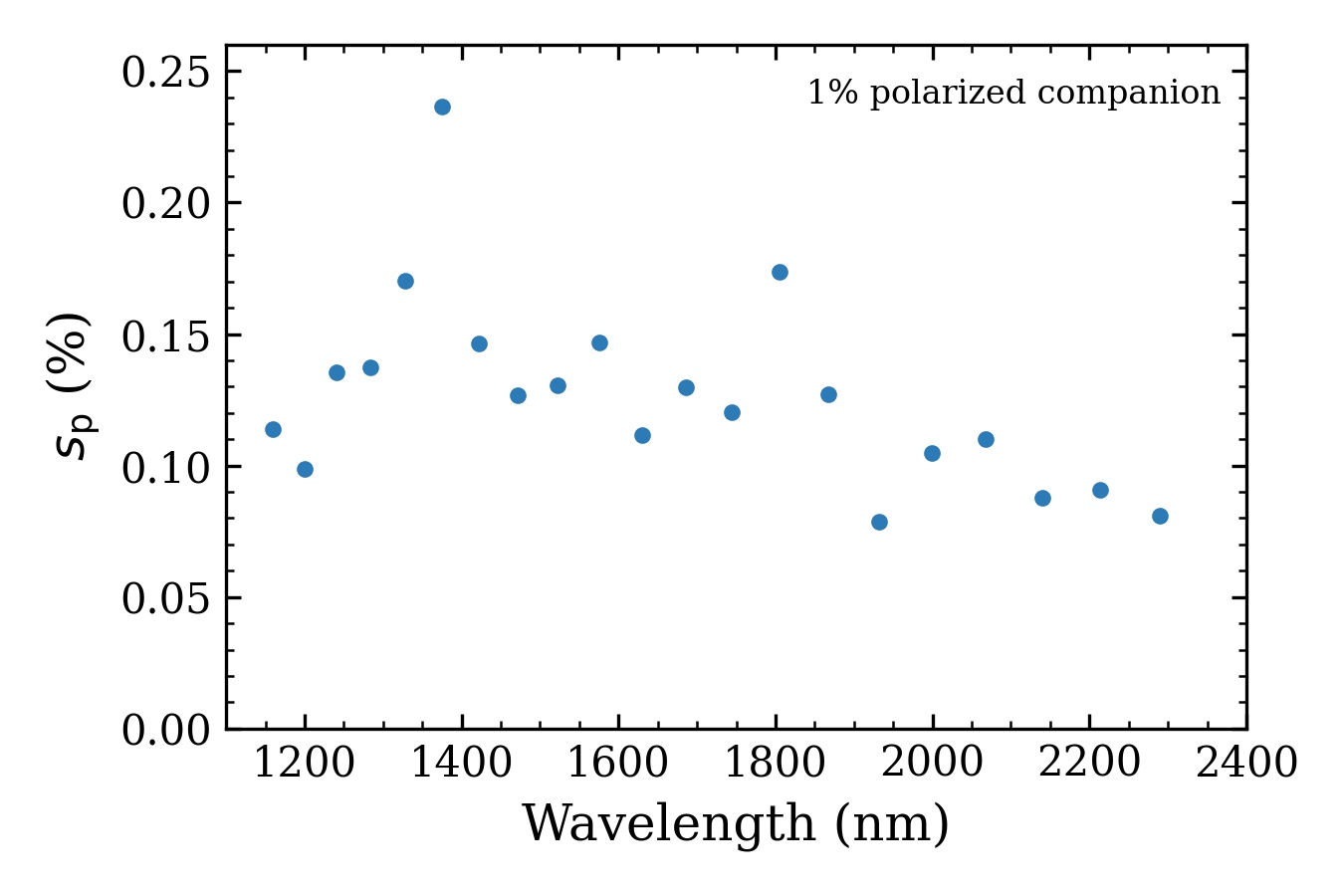}
         \caption{}
         \label{fig:dolp_acc_1_percent}
     \end{subfigure}
     \hfill
     \begin{subfigure}[b]{0.49\textwidth}
         \centering
         \includegraphics[width=\textwidth]{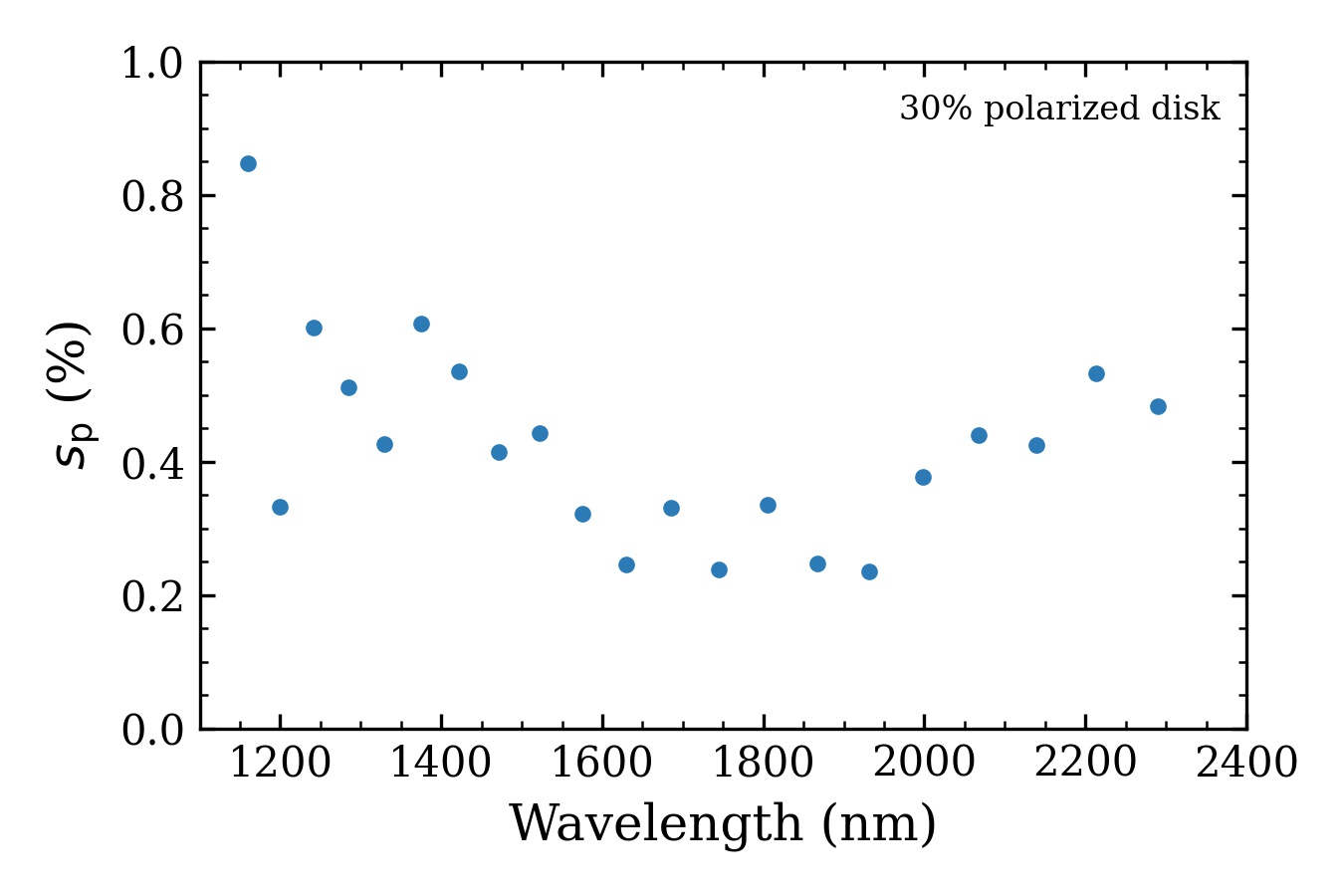}
         \caption{}
     \end{subfigure}
     \hfill
     \begin{subfigure}[b]{0.49\textwidth}
         \centering
         \includegraphics[width=\textwidth]{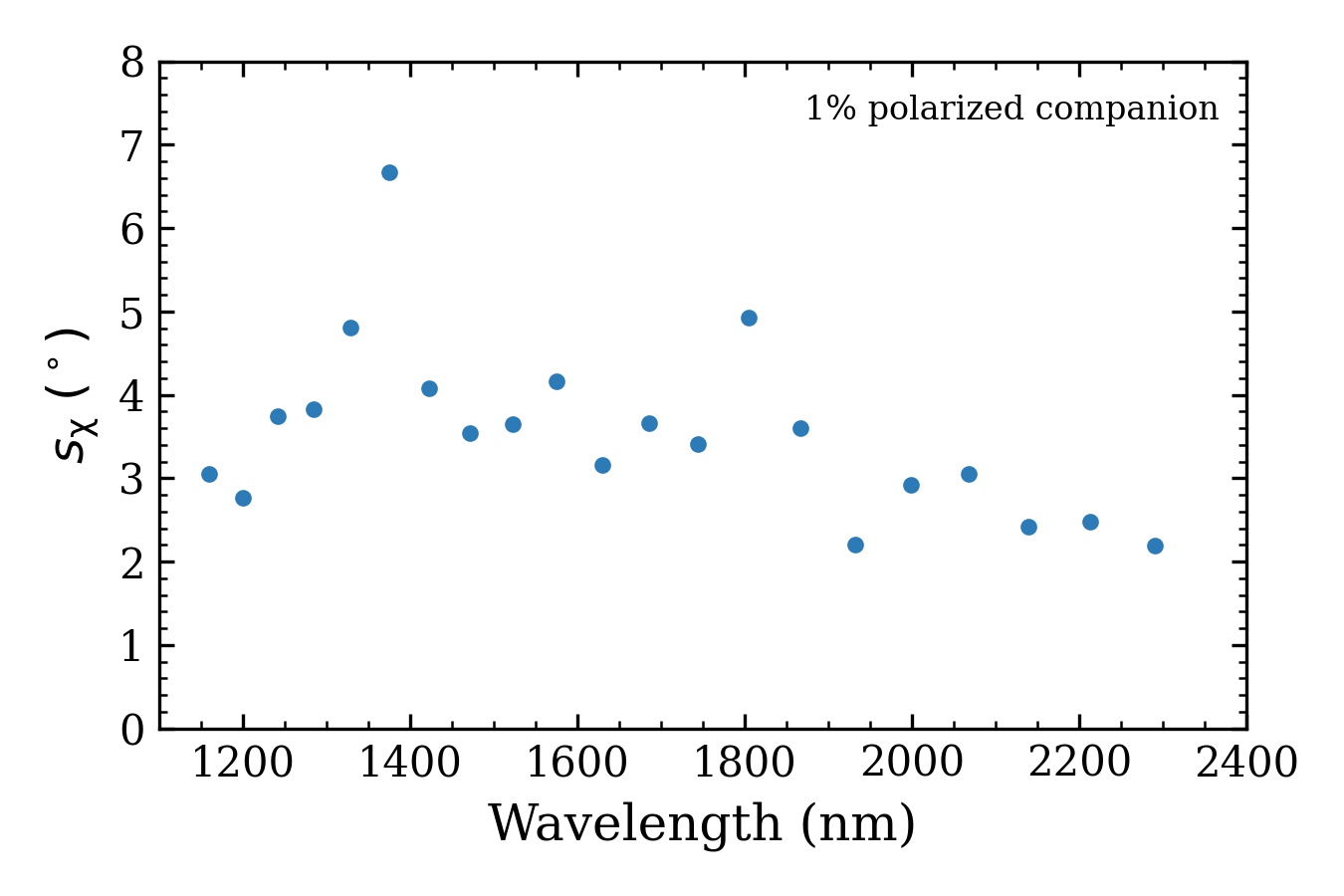}
         \caption{}
     \end{subfigure}
     \begin{subfigure}[b]{0.49\textwidth}
         \centering
         \includegraphics[width=\textwidth]{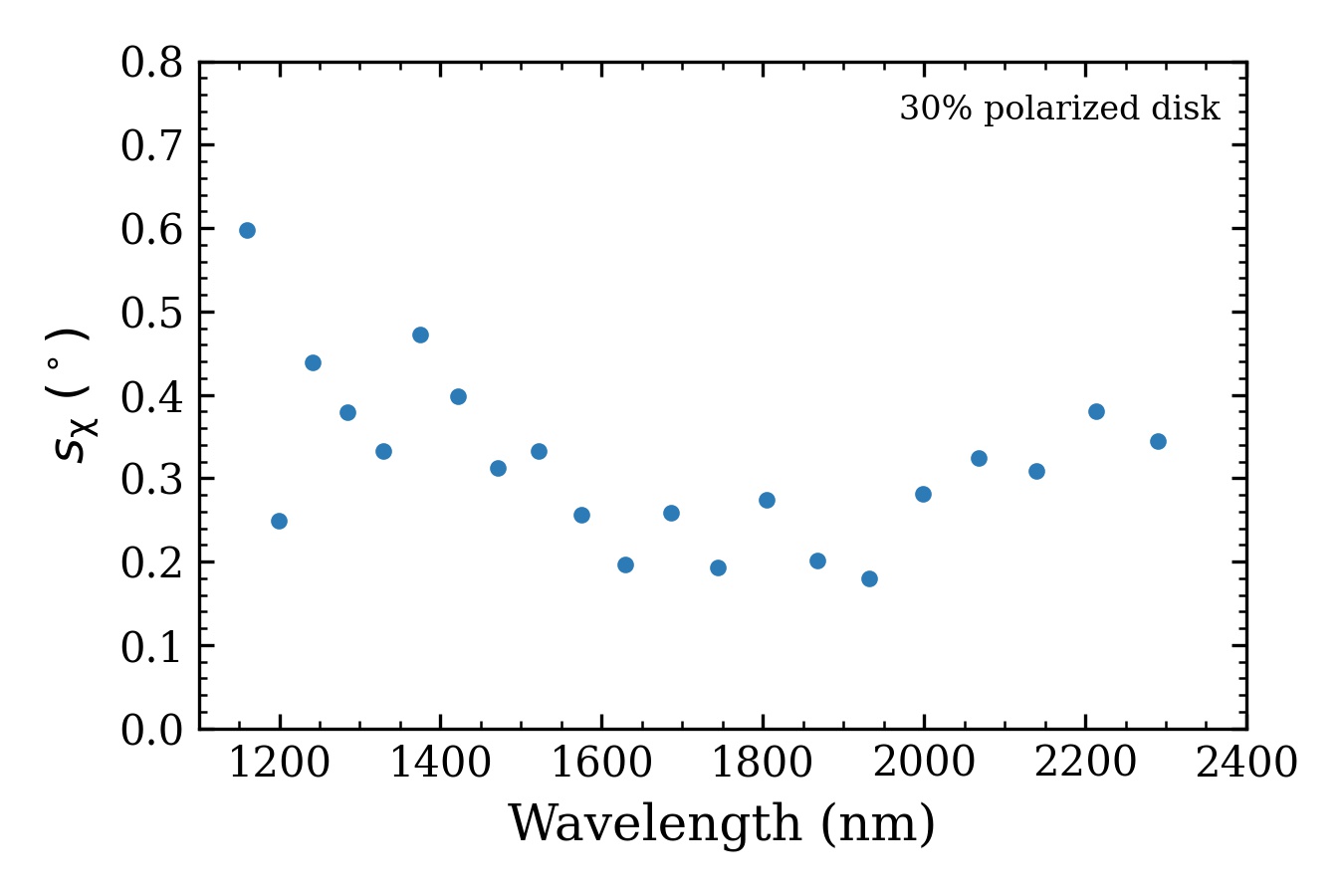}
         \caption{}
     \end{subfigure}
     \vspace{0.25cm}
        \caption{Polarimetric accuracies for measuring the degree of linear polarization, $P$, and the angle of linear polarization, $\chi$, for a $1\%$ polarized companion (a, c) and a $30\%$ polarized circumstellar disk (b, d). Recall that the goal of the work is to achieve $s_\mathrm{P} < 0.1 \%$ for all wavelength bins when observing a 1\% polarized companion.}
        \label{fig:polarimatric_accuracies}
\end{figure}

We can now calculate the total polarimetric accuracy in the Stokes parameters $Q$ and $U$ and use those to calculate the polarimetric accuracy on $P$ and $\chi$, $s_\mathrm{P}$ and $s_{\upchi}$. We calculate $s_\mathrm{P}$ and $s_{\upchi}$ for a 1\% polarized companion and a 30\% polarized disks for 21 wavelength bins. Recall that we excluded the longest wavelength bin of the unpolarized star observations (see \cref{sec:on_sky_results}). \Cref{fig:polarimatric_accuracies} shows the polarimetric accuracies. Because the calculated values depend on the angle of polarization, we show the worst case in the figures. For sources with low $P$ (a few percent), the accuracy is dominated by $s_\mathrm{abs}$. For sources with a higher $P$ (a few tens of percent), $s_\mathrm{rel}$ becomes more important. The accuracy on $\chi$ increases with increasing $P$, because the $Q$ and $U$ components of the light are measured with a higher accuracy. On the other hand, for sources with a very low $P$ (${<}0.1\%$) the error on the measurement of $\chi$ can become as high as a few tens of degrees.

In this work, we aimed for a polarimetric accuracy ${<}0.1\%$ on the degree of linear polarization for source with $P = 1\%$. \Cref{fig:dolp_acc_1_percent} shows that this is not achieved for all wavelength bins with the current model and data set. The limiting factor for the accuracy of our model is the absolute polarimetric accuracy, which is determined by the observations of the unpolarized star. This accuracy is primarily limited by the varying seeing during the observations and the saturation of the frames. To improve the accuracy of the model, higher-quality on-sky observations of an unpolarized standard star are needed.

%% file: discussion.tex
In this section, we discuss how we could improve the model in the future. First, we discuss the benefits of unpolarized internal source measurements in \cref{sec:unpolarized_internal_source_measurements}. Second, we propose improved observation strategies for on-sky calibration measurements in \cref{sec:future_calibration_measurements}. Third, in \cref{sec:verify_model_polarized_star}, we discuss how we can validate the model with observations of a circumstellar disk.

\subsection{Unpolarized Internal Source Measurements}\label{sec:unpolarized_internal_source_measurements}
The calibration measurements for the components downstream of the telescope are carried out by injecting nearly 100\% polarized light. Using these measurements, only the crosstalk of the HWP and derotator can be calibrated. To calibrate the IP of the HWP and the derotator, unpolarized light needs to be injected into the system by taking internal source measurements without using the calibration polarizer (see \cref{fig:optical_path}). 
However, the mirror of the calibration system is expected to produce IP as well. Similarly to Ref.~\citenum{vanHolstein2020IRDIS}, we can include the IP of the mirror of the calibration system in our model as a nuisance parameter. 

\subsection{Calibration Measurements of an Unpolarized Star}\label{sec:future_calibration_measurements}
As discussed in \cref{sec:on_sky_results}, the observations of HD 140667 suffered from varying seeing, and consequently, many of the frames are saturated or suffered from non-linearity. To accurately calibrate for the telescope diattenuation we propose to obtain new calibration observations of an unpolarized standard star. The observations have to be carried out such that the detector is not saturated. This can be achieved by observing at a night with stable and good seeing conditions. Furthermore, we can observe a fainter target than HD 140667. HD 140667 is relatively bright with a J-band magnitude of 6.4\cite{skiff2014vizier}. This resulted in the integration times being short, which makes it difficult to adjust the integration time for varying seeing conditions. A fainter target can be integrated longer, which also gives the AO system more time to average out. Also, to improve the accuracy of the measurement of the double sum it is important that the entire PSF of the source fits within the FOV of the detector. This will enable us to make a better estimation of the background in the frames. Lastly, the accuracy of the fit of the normalized Stokes parameters of the unpolarized star can be improved by obtaining an extra set of observations at low altitudes. With this additional set of observations, we will be able to better constrain the model and reach in all wavelength bins the polarimetric accuracy of ${<}0.1\%$ that we aimed for.

When calculating the normalized Stokes parameters, frames obtained with the HWP rotated over $45^\circ$ are combined. It is important that the conditions at which these frames are taken are similar. The observations used in this work are taken with the HWP rotating from $0^\circ$ to $67.5^\circ$ in steps of $22.5^\circ$. In terms of Stokes parameters, we observed in the order $Q^+$, $U^+$, $Q^-$, $U^-$. To minimize the time between the $+$ and $-$ measurements we propose to carry out future calibration observations with HWP orientations in the order $0^\circ$, $45^\circ$, $22.5^\circ$, $67.5^\circ$. This HWP-rotation method improves the accuracy of science observation as well. 


\subsection{Verifying the Model with Observations of a Circumstellar Disk}\label{sec:verify_model_polarized_star}
To validate the model developed in this work, we can apply the model to observations of a circumstellar disk for which the angle of linear polarization is well known, for example a face-on-viewed disk such as that of TW~Hya \cite{van2017three}. Before we can use the model, we need to add a model-based correction method to the CHARIS data-reduction pipeline, similar to Ref.~\citenum{vanHolstein2020IRDIS}, which corrects for IP and crosstalk. This data-reduction pipeline then has to be applied to the observations of the circumstellar disk to see whether we are able to obtain the correct $\chi$, and perhaps also $P$, of this system.

%% file: conclusion.tex
In this work, we have developed a detailed Mueller matrix model describing the instrumental polarization effects of the spectropolarimetric mode of SCExAO-CHARIS. We have used physical models to model the retardance of the image derotator (K-mirror) and half-wave plate (HWP) as well as the diattenuation of the telescope fold mirror M3. With these physical models, we are able to calibrate the crosstalk and instrumental polarization (IP) produced by the optical components in the instrument for all wavelength bins simultaneously by fitting a limited number of model parameters. The parameters describing the physical models are fitted to measurements obtained with an internal calibration source, which produces nearly $100\%$ linearly polarized light, and on-sky measurements of an unpolarized polarization standard star. 

The crosstalk of the system is strongly wavelength dependent and predominantly originates from the derotator. As a result of the crosstalk, the polarimetric efficiency can be as low as a few percent. From the polarized internal source measurements, we have seen that the crosstalk is strongest at $\lambda \approx 1600$ nm. Furthermore, the crosstalk is a function of the rotation angle of the derotator and strongest at $\theta_\mathrm{der} = 45^\circ$ ($\theta_\mathrm{der} = 135^\circ$). 

The IP is mainly caused by the telescope fold mirror (M3). The IP is a function of wavelength and increases for decreasing wavelength, reaching a maximum of ${\sim} 1\%$. The IP is independent of the altitude angle of the telescope. The physical model for M3 shows that even though the fitted refractive index of the mirror is close to that of a perfect silver mirror, the IP is significantly different. This clearly shows that calibration measurements are always necessary for the accurate calibration of IP.

The relative accuracy of the model ranges from $0.2\%$ to $2.5\%$ and the absolute accuracy from $0.16\%$ to $0.23\%$, for the different wavelength bins. These achieved accuracies will enable us to obtain a polarimetric accuracy range from $0.08\%$ to $0.24\%$ on observation of a $1\%$ polarized substellar companion. The goal we have set for this research is to obtain a polarimetric accuracy of ${<}0.1\%$ in all wavelength bins for such a target. Due to unstable seeing conditions during the observations of the unpolarized star and saturated data, we have not been able to reach this goal. Higher-quality observations of an unpolarized target are needed to reach a polarimetric accuracy of ${<}0.1\%$ in all wavelength bins.

With the model we developed, the crosstalk and IP of the system are known in advance. The implementation of a model-based correction of IP and crosstalk will enable us to obtain highly accurate measurements of the polarized intensity and angle of linear polarization. Our model will improve the accuracy of the spectropolarimetric measurements of SCExAO-CHARIS, which will enable unique spectropolarimetric observations of circumstellar disks and substellar companions.

%% file: acknowledgments.tex
The authors wish to recognize and acknowledge the very significant cultural role and reverence that the summit of Maunakea has always had within the indigenous Hawaiian community. We are very fortunate to have the opportunity to conduct observations from this mountain. The research of S.P. Bos and F. Snik leading to these results has received funding from the European Research Council under ERC Starting Grant agreement 678194 (FALCONER). The development of SCExAO was supported by the Japan Society for the Promotion of Science (Grant-in-Aid for Research \#23340051, \#26220704, \#23103002, \#19H00703 \& \#19H00695), the Astrobiology Center of the National Institutes of Natural Sciences, Japan, the Mt Cuba Foundation and the director's contingency fund at Subaru Telescope.

%% file: appendix_diattenuation_of_derotator.tex
\begin{figure}[h]
    \centering
    \includegraphics[width=0.49\textwidth]{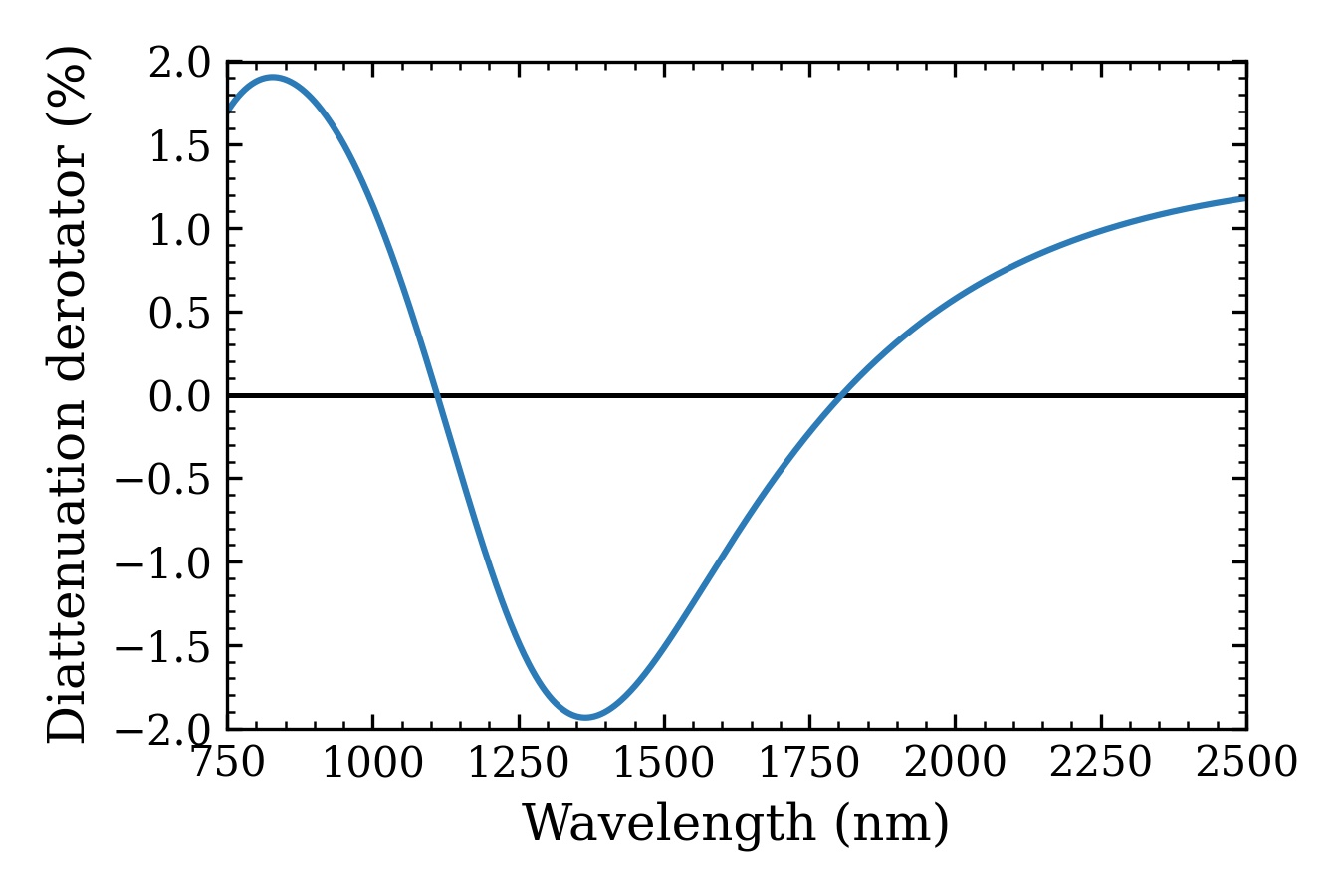}
    \caption{Estimated diattenuation of the derotator as a function of wavelength.}
    \label{fig:diatt_der}
\end{figure}

In this appendix, we look at the diattenuation of the derotator mirrors. Using the nearly $100\%$ linearly polarized internal source measurements, we can only directly measure the crosstalk created by the retardance of the derotator mirrors. As discussed in \ref{sec:derotator_model}, we use a physical model to calculate the Fresnel coefficients of the multi-layer reflection of the three derotator mirrors. Using the fitted film width, $d$, we can calculate the Fresnel coefficients and use them to calculate the diattenuation of the derotator. \Cref{fig:diatt_der} shows the diattenuation of the derotator derived from the multi-layer reflection model using the fitted value of $d$ (see \cref{tab:instrument_model_parameters}). The figure shows that the diattenuation of the derotator is strongly wavelength dependent and ranges between $2\%$ and ${-}2\%$. Using the nearly $100\%$ linearly polarized internal source measurements we cannot validate this results. However, the result shows that the diattenuation of the derotator can be significant. Additional unpolarized internal source measurements, as discussed in \cref{sec:unpolarized_internal_source_measurements}, are necessary to calibrate for the diattenuation of the derotator.

